\documentclass[a4paper]{cas-sc}

\usepackage[authoryear,longnamesfirst]{natbib}

\def\tsc#1{\csdef{#1}{\textsc{\lowercase{#1}}\xspace}}
\tsc{WGM}
\tsc{QE}

\usepackage{lipsum}
\usepackage{amssymb}
\usepackage{amsmath}
\usepackage{algorithm}
\usepackage{algpseudocode}
\usepackage{booktabs}
\usepackage{amsthm}
\usepackage{xcolor}
\usepackage{tikz}
\usepackage{subcaption}
\usepackage{amsthm}
\usepackage{caption}
\usepackage{enumitem}
\usepackage{listings}

\lstdefinestyle{matlab-style}{
  language=Matlab,
  basicstyle=\ttfamily\small,
  keywordstyle=\color{blue},
  stringstyle=\color{orange},
  commentstyle=\color{gray},
  numbers=left,
  numberstyle=\tiny\color{gray},
  stepnumber=1,
  numbersep=10pt,
  backgroundcolor=\color{white},
  frame=single,
  breaklines=true,
  captionpos=b,
  tabsize=4,
  showstringspaces=false
}

\newdefinition{definition}{Definition}
\newdefinition{parametrization}{Parametrization}
\newdefinition{lemma}{Lemma}

\newtheorem{theorem}{Theorem}
\theoremstyle{remark}
\newtheorem{remark}{Remark}
\newtheorem{algorithm2}{Algorithm}
\newcommand{\hop}{\mathsf{H}}

\definecolor{dblue}{rgb}{0 0 0.7}
\definecolor{red}{rgb}{1 0 0}
\definecolor{MatlabBlue}{rgb}{0, 0.4470, 0.7410}
\definecolor{MatlabRed}{rgb}{0.6350, 0.0780, 0.1840}
\definecolor{MatlabOrange}{rgb}{0.8500, 0.3250, 0.0980}
\definecolor{MatlabYellow}{rgb}{0.9290, 0.6940, 0.1250}
\definecolor{MatlabPurple}{rgb}{0.4940, 0.1840, 0.5560}
\definecolor{MatlabGreen}{rgb}{0.4660, 0.6740, 0.1880}
\definecolor{MatlabBabyBlue}{rgb}{0.3010, 0.7450, 0.9330}
\definecolor{MatlabGray}{rgb}{0.5, 0.5, 0.5}
\definecolor{MatlabLightGray}{rgb}{0.75, 0.75, 0.75}
\definecolor{MatlabBlack}{rgb}{0, 0, 0}
\definecolor{MatlabLightGray4}{rgb}{0.875, 0.875, 0.875}
\definecolor{MatlabLightGray3}{rgb}{0.85, 0.85, 0.85}
\definecolor{MatlabLightGray2}{rgb}{0.775, 0.775, 0.775}
\definecolor{MatlabLightGray1}{rgb}{0.7, 0.7, 0.7}
\definecolor{MatlabGray20}{rgb}{0.2, 0.2, 0.2}
\definecolor{MatlabGray30}{rgb}{0.3, 0.3, 0.3}
\definecolor{MatlabGray40}{rgb}{0.4, 0.4, 0.4}
\definecolor{MatlabGray50}{rgb}{0.5, 0.5, 0.5}
\definecolor{MatlabGray60}{rgb}{0.6, 0.6, 0.6}
\definecolor{MatlabGray70}{rgb}{0.7, 0.7, 0.7}
\definecolor{MatlabGray80}{rgb}{0.8, 0.8, 0.8}
\definecolor{MatlabGray85}{rgb}{0.85, 0.85, 0.85}
\definecolor{MatlabGray90}{rgb}{0.9, 0.9, 0.9}

\newcommand{\tikzline}[1]{(\protect\tikz[baseline=-0.6ex,x=1pt,y=1pt,line width=0.4mm]{\protect\draw[#1] (0,0) -- (10,0);})}

\newcommand{\tikzmarkline}[1]{(\protect\tikz[baseline=-0.6ex,x=1pt,y=1pt]{\protect\draw[#1] (0,0) -- (10,0);\protect\draw[color=#1, thick] (5,0) circle (2pt)})}

\usepackage{eso-pic}
\AddToShipoutPictureBG*{%
  \AtPageUpperLeft{%
    \setlength\unitlength{1in}%
    \hspace*{\dimexpr0.5\paperwidth\relax}%
    \makebox(0,-0.9)[c]{%
      \begin{tabular}{c}
        Maarten van der Hulst \emph{et al.}, \\
        \emph{Structured identification of multivariable modal systems}, \\
        uploaded to arXiv \today \\
      \end{tabular}%
    }%
  }%
}

\AddToShipoutPictureFG*{%
  \AtPageUpperLeft{%
    \color{white}\rule{\paperwidth}{1.2cm}%
  }%
  \AtPageLowerLeft{%
    \color{white}\rule{\paperwidth}{1.4cm}%
  }%
}

\makeatletter
\def\pprintMaketitle{%
    \clearpage
    \thispagestyle{empty}%
    \begin{center}%
      {\Large\@title\par}%
      \vskip 1em%
      {\normalsize
        \lineskip .5em%
        \begin{tabular}[t]{c}%
          \@author
        \end{tabular}\par}%
      \vskip 1em%
      {\small \@date}%
    \end{center}%
    \par
    \vskip 1.5em
}
\makeatother


 \usepackage{float}
\usepackage{placeins}

\begin{document}
\let\WriteBookmarks\relax
\def\floatpagepagefraction{1}
\def\textpagefraction{.001}

% \begin{frontmatter} % <- not needed with cas-sc

\title[mode=title]{Structured identification of multivariable modal systems}

\author[1]{M. van der Hulst} 
\author[1]{R. A. Gonz\'alez}
\author[1]{K. Classens}
\author[1]{P. Tacx}
\author[2]{N. Dirkx}
\author[2]{J. van de Wijdeven}
\author[1,3]{T. Oomen}

\affiliation[1]{organization={Deptartment of Mechanical Engineering, Eindhoven University of Technology},
                country={The Netherlands}}

\affiliation[2]{organization={ASML},
                city={Veldhoven},
                country={The Netherlands}}

\affiliation[3]{organization={Delft Center for Systems and Control, Delft University of Technology},
                country={The Netherlands}}

\tnotetext[1]{This project is funded by Holland High Tech | TKI HSTM via the PPP Innovation Scheme (PPP-I) for public-private partnerships.}

%\shorttitle{}
%\shortauthors{M. van der Hulst et al.,}

\begin{abstract}
Physically interpretable models are essential for next-generation industrial systems, as these representations enable effective control, support design validation, and provide a foundation for monitoring strategies. The aim of this paper is to develop a system identification framework for estimating modal models of complex multivariable mechanical systems from frequency response data. To achieve this, a two-step structured identification algorithm is presented, where an additive model is first estimated using a refined instrumental variable method and subsequently projected onto a modal form. The developed identification method provides accurate, physically-relevant, minimal-order models, for both generally-damped and proportionally damped modal systems. The effectiveness of the proposed method is demonstrated through experimental validation on a prototype wafer-stage system, which features a large number of spatially distributed actuators and sensors and exhibits complex flexible dynamics.
\end{abstract}


\maketitle

\section{Introduction}
Increasingly stringent performance requirements of next-generation industrial systems require data-driven models to integrate prior knowledge via interpretable physical principles. Embedding such principles improves model accuracy, increases interpretability, and typically reduces model complexity \cite{Soderstrom2001SystemIdentification}. These characteristics are crucial not only for advanced multivariable control design methods \cite{VanDeWal2002MultivariableMotion, VanHerpen2014ExploitingControl} but also for mechanical system design, model validation, machine monitoring techniques, and simulation studies \cite{Schmidt2011TheMechatronics, Heertjes2020ControlDevelopments,Steinbuch2022MotionLaw, Classens2023FromReconfiguration}. 

The use of modal model structures for modeling mechanical systems enables efficient representations that capture the underlying physics~\cite{M.M.Silva1997ModalAnalysis, Ewins2000ModalTesting}. In such models, the dynamics are expressed as a combination of independently evolving dynamic modes, each associated with a resonance behavior of the system. Key advantages include that modal models directly realize the system’s minimal order, which is critical for applications in control system design~\cite{Skogestad2006MultivariableDesign}, while still offering clear physical interpretation and computational tractability. As a result, modal forms are widely used in applications such as mechatronic positioning systems \cite{Verkerk2012ModalPositions,VanHerpen2014ExploitingControl, Voorhoeve2021IdentifyingStage}, fault detection  \cite{Classens2022FaultResonances, Classens2025RecursiveSystems}, finite element model updating \cite{DaSilva2008DesignOptimization, Dorosti2018IterativeApproach}, vibrational analysis of aerospace systems \cite{Vayssettes2015NewStructures}, adaptive optics \cite{ Tacx2024IdentificationSystems} and civil engineering applications \citep{Zhang2019VariationalEngineering}. 

Reliably estimating modal input–output models remains challenging, primarily due to the structural constraints of the modal form and the complexity of industrial systems. A defining characteristic of modal decompositions is the requirement that the numerator matrix associated with each mode satisfies a low-rank constraint \cite{DeKraker2004ADynamics}. Traditional parametric system identification methods do not incorporate such constraints \cite{Ljung1999SystemUser}, leading to increased model orders \cite{Gustavsen2004ADomain} and reduced interpretability. Enforcing exact rank constraints is in general NP-hard \cite{Fazel2002MatrixApplications}, requiring specialized optimization methods or carefully designed relaxations for tractability. Another structural aspect is that modal systems are naturally expressed in additive form, that is, as a sum of low-order transfer matrices~\cite{Gawronski2004AdvancedStructures}. Most identification methods, however, are tailored to single unfactored transfer models and are therefore less suitable for the estimation of modal systems \cite{Gilson2018AIdentification}. Moreover, industrial systems often feature many inputs and outputs and exhibit complex dynamics such as closely spaced and lightly damped flexible modes. These characteristics lead to poor numerical conditioning~\cite{Voorhoeve2014OnMethods,VanHerpen2014OptimallyIdentification} and high-dimensional estimation problems, thereby complicating the modeling task.

Despite extensive work on modal identification, relatively few methods explicitly enforce the rank constraint inherent to modal decompositions. Traditional approaches include single-step methods, which approximate the original nonlinear least-squares problem by a linear surrogate \cite{ Peeters2004TheEstimation, Peeters2005PolyMax:Analysis}. Improved estimation accuracy has been achieved through iterative multi-step methods that directly minimize the nonlinear cost function, using gradient-descent optimization \cite{Bayard1994High-orderResults}, pseudo-linear regression techniques \cite{Sanathanan1961TransferPolynomials, Blom2010MultivariableRegression} or vector fitting procedures \cite{Semlyen1999RationalFitting,  Civera2021ExperimentalMethod}. These approaches have been applied in a range of multi-input multi-output (MIMO) modal identification methods \cite{Verboven2005MultivariableEstimation, Vayssettes2014Frequency-domainTests, Reynders2012SystemComparison, Voorhoeve2021IdentifyingStage}. However, single-step methods generally lack statistical optimality guarantees, while many multi-step approaches either neglect the rank constraint or rely on singular value decompositions, which are typically effective only for systems with sufficiently spaced modes.

While some efforts have been made to incorporate the rank constraint, the range of available methods remains limited. Post-processing strategies, such as data refitting, transform an initially estimated model into modal form and then re-estimate the parameters~\cite{Voorhoeve2021IdentifyingStage, Gustavsen2004ADomain}. In contrast, integrated approaches aim to enforce the structure directly, either by incorporating modal properties into the parameterization~\cite{Voorhoeve2016IdenticationConsiderations, Vayssettes2015StructuredOptimisation} or by using parameter projection techniques~\cite{Mercere2014IdentificationOne, Yu2018IdentificationModels}. While data refitting can yield accurate modal estimates, it can become computationally demanding, particularly in large-scale MIMO settings. Parameter projection methods offer a promising alternative, but their development has so far been largely limited to state-space model estimation.

Although many parametric modal identification techniques are available, {there is currently no systematic and computationally efficient framework that explicitly enforces the rank constraints intrinsic to modal decompositions,  exploits the additive structure of modal models in the model structure, and remains applicable to large-scale industrial systems with closely spaced modes.} This paper introduces a novel two-stage identification strategy for estimating MIMO input-output models in modal form. In the first stage, an additive model is estimated using a refined instrumental variable method \cite{Young1980RefinedAnalysis}, which is subsequently projected onto a modal form using the indirect prediction error method (IPEM) \cite{Soderstrom1991AnIdentification}. The framework accommodates both proportional damping and general viscous damping models \cite{DeKraker2004ADynamics}, as these formulations are commonly encountered in industrial applications \cite{Verbaan2015RobustStages}. The main contributions of this paper are:
\begin{enumerate}[label=C\arabic*:, leftmargin=*, align=left]
  \item A general framework is introduced for estimating continuous-time MIMO models in additive transfer-function form using a frequency-domain refined instrumental variable method.
  \item A structured identification framework is introduced that projects additive models onto the modal form, accommodating both proportional and general viscous damping model parameterizations.
\end{enumerate}

A preliminary version addressing C1 was presented in \cite{vanderHulst2025FrequencyStage}. In addition, the two-stage strategy for structured model estimation may be interpreted as the frequency-domain counterpart of the time-domain method introduced in \cite{Gonzalez2025StatisticallyIdentification}, which includes an extensive statistical analysis. In this paper, we extend these results to a broader class of damping models and provide a comprehensive experimental validation on a complex industrial system.

The remainder of this paper is organized as follows. Section 2 introduces the modal modeling framework for flexible mechanical systems, and provides a concise problem formulation. Section 3 formulates the two‐stage identification problem, with the solution procedures for the first stage provided in Section 4 and for the second stage in Section 5. Experimental validation is described in Section 6, and finally, Section 7 presents the conclusions.

\textit{Notation:}  Scalars, vectors, matrices and sets are written as $a$, $\mathbf{a}$, $\mathbf{A}$ and $\mathcal{A}$, respectively. The imaginary unit is denoted by $j$ and satisfies $ j^2 = -1 $. For $\mathbf{z}\in\mathbb{C}^n$, the operation $\Re\{\mathbf{z}\}$ returns the real part, $\Im\{\mathbf{z}\}$ returns the imaginary part and $\bar{\mathbf{z}}$ returns the complex conjugate of the complex vector $\mathbf{z}$. The identity matrix of size $n$ is denoted by $\mathbf{I}_{n}$. For a matrix $ \mathbf{A} $, its transpose is written as $ \mathbf{A}^{\top} $, and its Hermitian (conjugate transpose) as $ \mathbf{A}^{\hop} $. If $ \mathbf{x} \in \mathbb{C}^n $ and $ \mathbf{Q} \in \mathbb{C}^{n \times n} $ is a Hermitian matrix, then the weighted 2-norm is given by $ \|\mathbf{x}\|_{\mathbf{Q}} = \sqrt{\mathbf{x}^{\hop} \mathbf{Q} \mathbf{x}} $. For $ \mathbf{X} = [\mathbf{x}_1, \ldots, \mathbf{x}_n] $, with $ \mathbf{x}_i \in \mathbb{C}^n $, the operation $ \operatorname{vec}(\mathbf{X}) = [\mathbf{x}^\top_1, \ldots, \mathbf{x}^\top_n]^\top $ restructures the matrix into a vector by stacking its columns.

\newpage

\section{Problem formulation} 
This section first introduces the considered modal modeling framework for describing mechanical systems and thereafter formulates the problem addressed in this paper.

\subsection{Modal modeling of mechanical systems}
The equations of motion for a spatially discretized linear-time invariant (LTI) mechanical system of order $2n$ with $ n_{\mathrm{u}} $ inputs and $ n_{\mathrm{y}} $ outputs are defined in terms of the \textit{generalized} coordinates $ \mathbf{q}(t) \in \mathbb{R}^{n} $ as follows
\begin{equation} \label{eq:  eom nodal coordinates}
\begin{aligned}
&\mathbf{M}\ddot{\mathbf{q}}(t) + \mathbf{D}\dot{\mathbf{q}}(t) + \mathbf{K}\mathbf{q}(t) = \mathbf{F}\mathbf{u}(t),\\
&\mathbf{y}(t) = \mathbf{Q}\mathbf{q}(t),
\end{aligned}
\end{equation}
{with $\mathbf{y}(t)\in \mathbb{R}^{n_\mathrm{y}}$ the system outputs, $\mathbf{u}(t)\in \mathbb{R}^{n_\mathrm{u}}$ the system inputs}, the mass matrix $\mathbf{M} \in \mathbb{R}^{n \times n}$ is positive definite, the viscous damping matrix is given by $\mathbf{D} \in \mathbb{R}^{n \times n }$, the stiffness matrix $\mathbf{K} \in \mathbb{R}^{n \times n}$ is positive semi-definite, $\mathbf{F}\in \mathbb{R}^{n \times n_{\mathrm{u}}}$ is the input matrix, and $\mathbf{Q}\in \mathbb{R}^{n_{\mathrm{y}}\times n }$ is the output  matrix. The coupled second-order differential equations are expressed in modal form, following standard approaches in vibration analysis \cite{Ewins2000ModalTesting,DeKraker2004ADynamics,Gawronski2004AdvancedStructures}. Two modal formulations are considered. {The first corresponds to generally viscously damped systems, while the second assumes proportional damping, a special case in which the damping matrix is proportional to the mass and stiffness matrices, resulting in mode-shape vectors that are real-valued and coincide with those of the undamped system \cite{Gawronski2004AdvancedStructures}. The choice between damping models depends on the system under consideration. While proportional damping provides a simpler and more interpretable model description due to the real-valued mode-shape vectors, it can be inadequate for systems involving more complex damping behavior, necessitating the use of the general form (see, e.g., \cite{Verbaan2015RobustStages}).} A complete derivation of the modal forms is provided in Appendix~A, and the main results are presented in the following subsections.

\subsubsection{General-viscously damped mechanical systems}
The general case is considered first. The dynamics are partitioned into $ n_{\mathrm{rbm}} $ rigid-body modes and $ n_{\mathrm{flex}} $ flexible modes{, where $n=n_{\mathrm{rbm}} + n_{\mathrm{flex}}$}. A rigid-body mode represents a motion of the system involving pure translation or rotation without internal deformation and is described by a second-order model with two poles at $s=0$. {While physical systems contain infinitely many dynamic modes, only the dominant modes within the excited and measured frequency range are estimated explicitly. Modes at substantially higher frequencies typically have only a minor effect in this range, and their combined contribution can be approximated by a residual static gain \cite{Gawronski2004AdvancedStructures}.} This term is omitted for notational clarity but is straightforwardly included if required. The transfer matrix corresponding to \eqref{eq: eom nodal coordinates}, expressed in modal form with general damping, is given~by
\begin{align} \label{eq: modal transfer first order}
\mathbf{P}_m(s) &= \sum_{j=1}^{n_{\mathrm{rbm}}} \frac{\mathbf{R}^{\mathrm{rbm}}_j}{s^2} 
+ \sum_{i=1}^{n_{\mathrm{flex}}}\frac{\mathbf{L}_i}{s - \lambda_i} + \frac{\bar{\mathbf{L}}_i}{s - \bar{\lambda}_i}, \\
\mathbf{L}_i &= \boldsymbol{\psi}_{l,i} \boldsymbol{\psi}_{r,i}^{\top}, \label{eq: resid gen} \\
\mathbf{R}^{\mathrm{rbm}}_j  &= \boldsymbol{\phi}_{l,j} \boldsymbol{\phi}_{r,j}^{\top},
\end{align}
where $ s \in \mathbb{C}$ is the Laplace variable, $ \mathbf{R}^{\mathrm{rbm}}_j \in \mathbb{R}^{n_{\mathrm{y}} \times n_{\mathrm{u}}} $ and $ \mathbf{L}_i \in \mathbb{C}^{n_{\mathrm{y}} \times n_{\mathrm{u}}} $ are rank-one modal residue matrices, $ \boldsymbol{\psi}_{l,i} \in \mathbb{C}^{n_{\mathrm{y}}} $ and $ \boldsymbol{\psi}_{r,i} \in \mathbb{C}^{n_{\mathrm{u}}} $ are the complex left and right mode-shape vectors of the flexible modes, $ \boldsymbol{\phi}_{l,j} \in \mathbb{R}^{n_{\mathrm{y}}} $ and $ \boldsymbol{\phi}_{r,j} \in \mathbb{R}^{n_{\mathrm{u}}} $ are the real left and right mode-shape vectors of the rigid-body modes, and $ \lambda_i \in \mathbb{C} $ is the eigenvalue, given by
\begin{equation} \label{eq: eigenvalue}
\lambda_i = -\zeta_i \omega_i + j \omega_i \sqrt{1 - \zeta_i^2},
\end{equation}
with $ \omega_i > 0 $ the natural frequency and $ \zeta_i > 0 $ the damping coefficient. Note that the decomposition of rank-one residue matrices \( \mathbf{L}_i = \boldsymbol{\psi}_{l,i} \boldsymbol{\psi}_{r,i}^\top \) is not unique, since any alternative factorization satisfies \( \mathbf{L}_i = \tilde{\boldsymbol{\psi}}_{l,i} \tilde{\boldsymbol{\psi}}_{r,i}^\top \), where \( \tilde{\boldsymbol{\psi}}_{l,i} = \alpha \boldsymbol{\psi}_{l,i} \) and \( \tilde{\boldsymbol{\psi}}_{r,i} = \alpha^{-1} \boldsymbol{\psi}_{r,i} \) for some nonzero scalar \( \alpha \in \mathbb{C}\)~\cite{Horn2013MatrixEdition}. This invariance allows for arbitrary scaling of the mode-shape vectors, which is typically resolved by applying an appropriate normalization \cite{DeKraker2004ADynamics}. To obtain a representation expressed in real-valued model parameters, the conjugated first-order terms in \eqref{eq: modal transfer first order} are combined, yielding
\begin{align} \label{eq: modal transfer second order}
\mathbf{P}_{m}(s) &= \sum_{j=1}^{n_{\mathrm{rbm}}} \frac{\mathbf{R}^{\mathrm{rbm}}_j}{s^2} 
+ \sum_{i=1}^{n_{\mathrm{flex}}} \frac{\mathbf{N}_{i,1} s + \mathbf{N}_{i,0}}{s^2 - (\lambda_i + \bar{\lambda}_i)s + \lambda_i \bar{\lambda}_i}, \\
\mathbf{N}_{i,1}  &= \mathbf{L}_i + \bar{\mathbf{L}}_i, \label{eq: term 1} \\
\mathbf{N}_{i,0}  &= -\bar{\lambda}_i \mathbf{L}_i - \lambda_i \bar{\mathbf{L}}_i,  \label{eq: term 2}
\end{align}
where  $ \mathbf{N}_{i,1} \in \mathbb{R}^{n_{\mathrm{y}} \times n_{\mathrm{u}}} $ and $ \mathbf{N}_{i,0}\in \mathbb{R}^{n_{\mathrm{y}} \times n_{\mathrm{u}}} $ are real-valued rank-two matrices. The parameters describing the general viscously damped modal system are jointly stored in the parameter vector
\begin{equation}\label{eq: modal parameter vector}
\boldsymbol{\rho} =
\begin{bmatrix}
\boldsymbol{\varrho}^{\top}_1 & \ldots & \boldsymbol{\varrho}^{\top}_{n_{\mathrm{rbm}}} &
\boldsymbol{\rho}^{\top}_1 & \ldots & \boldsymbol{\rho}^{\top}_{n_{\mathrm{flex}}}
\end{bmatrix}^{\top}.
\end{equation}
where {the vector \(\boldsymbol{\varrho_j}\) denotes the parameter vector of the \(j\)th rigid-body mode, and \(\boldsymbol{\rho}_i\) denotes the parameter vector of the \(i\)th flexible dynamic mode, which are both given by
}  
\begin{align}
\boldsymbol{\varrho}_j &= 
\begin{bmatrix}
\boldsymbol{\phi}_{l,j}^{\top} & \boldsymbol{\phi}_{r,j}^{\top}
\end{bmatrix}^{\top},\quad \boldsymbol{\rho}_i = 
\begin{bmatrix}
\Re\lambda_i & \Im\lambda_i &
\Re\,\boldsymbol{\psi}_{l,i}^{\top} & \Im\,\boldsymbol{\psi}_{l,i}^{\top} &
\Re\,\boldsymbol{\psi}_{r,i}^{\top} & \Im\,\boldsymbol{\psi}_{r,i}^{\top}
\end{bmatrix}^{\top}.
\end{align}

\subsubsection{Proportionally damped mechanical systems}
In contrast to the general case, proportional damping assumes that the damping matrix can be expressed as a linear combination of the mass and stiffness matrices~\cite{DeKraker2004ADynamics}. Consequently, the mode shapes coincide with the mode shapes of the undamped system and remain real-valued, whereas in the general case they are typically complex-valued. Under proportional damping, the transfer matrix in~\eqref{eq: modal transfer second order} simplifies to
\begin{align} \label{eq: modal transfer second order propportional}
\mathbf{P}_m(s) &= \sum_{j=1}^{n_{\mathrm{rbm}}}\frac{\mathbf{R}^{\mathrm{rbm}}_j}{s^2} + \sum_{i=1}^{n_{\mathrm{flex}}}  \frac{\mathbf{R}^{\mathrm{flex}}_i}{s^2 + 2 \zeta_i \omega_i s + \omega_i^2}, \\
\mathbf{R}^{\mathrm{flex}}_i &= \boldsymbol{\phi}_{l,i} \boldsymbol{\phi}_{r,i}^\top, \label{eq: resid prop}
\end{align}
where $ \boldsymbol{\phi}_{l,i} \in \mathbb{R}^{n_{\mathrm{y}}} $ and $\boldsymbol{\phi}_{r,i} \in \mathbb{R}^{n_{\mathrm{u}}} $. The parameter vector describing proportionally damped modal systems is given by \eqref{eq: modal parameter vector} but with $ \boldsymbol{\rho}_i $ for $ i = 1,\ldots,n_{\mathrm{flex}}$ for the flexible dynamic modes replaced by
\begin{equation}
\boldsymbol{\rho}_i =
\begin{bmatrix}
\omega_i & \zeta_i & \boldsymbol{\phi}_{l,i}^{\top} & \boldsymbol{\phi}_{r,i}^{\top}
\end{bmatrix}^{\top}.
\end{equation}

\subsection{Problem statement}
This paper addresses the estimation of input–output models in modal form, as defined in~\eqref{eq: modal parameter vector}, for both proportionally damped and generally viscously damped mechanical systems using experimental data. The identification problem is structured, since the numerators are constrained to the dyadic product of the mode-shape vectors. A frequency-domain identification approach is adopted, in which the objective is to estimate a model that best fits a measured Frequency Response Function (FRF) of the system. To obtain continuous-time models from FRF data, the setup outlined in \cite[Chapter~13]{Pintelon2012SystemIdentification} is used. {The noisy FRF dataset is denoted by \( \{ \mathbf{G}(\omega_k) \}_{k=1}^{N} \), with \( N \) the number of frequency points and \( \mathbf{G}(\omega_k) \in \mathbb{C}^{n_y \times n_u} \) the response at frequency $k$, and forms the basis for subsequent parametric identification.}

\newpage
\section{Two-stage modal identification approach}
This section introduces the identification strategy used to estimate modal systems from frequency response data. The modal model structure requires a structured identification approach, since the numerators formed by the dyadic products of the mode-shape vectors impose a rank-one constraint on the residue matrices. Enforcing rank constraints is generally NP-hard \cite{Fazel2002MatrixApplications}. To address this challenge, a two-stage identification approach is adopted. In the first stage, an additive model is estimated with freely parameterized numerator matrices. In the second stage, the estimated additive model is projected onto the subspace of models satisfying the rank constraint to obtain a model consistent with the modal form. In the following subsections, the identification problems for the first and second stages are formally defined.

\subsection{Stage 1: Identification of additive MIMO systems}
This subsection introduces the additive model structure and formulates the additive model identification problem. Consider the LTI plant model in additive form
\begin{equation} \label{eq: FREQ - true plant}
\mathbf{P}(s,\boldsymbol{\beta}) = \sum_{i=1}^K\mathbf{P}_i(s,\boldsymbol{\theta}_i),
\end{equation}
where each submodel $\mathbf{P}_i$ is an $n_{\mathrm{y}}\times n_{\mathrm{u}}$ transfer matrix, $K$ the number of submodels and $\boldsymbol{\beta}$ and $\boldsymbol{\theta}_i$ the joint and submodel parameter vector, respectively.  Each submodel $\mathbf{P}_i(s,\boldsymbol{\theta}_i)$ is parametrized according to
\begin{equation}
\mathbf{P}_i(s,\boldsymbol{\theta_i}) = \frac{1}{ s^{\ell_i} A_i(s)}\mathbf{B}_i(s),
\end{equation} 
where at most one submodel may include $\ell_i>0$ poles at the origin. The scalar denominator polynomial $A_i(s)$ and the matrix numerator polynomial $\mathbf{B}_i(s)$ are such that no complex number $z$ simultaneously satisfies $A_i(z) = 0$ and $ \mathbf{B}_i(z) = \mathbf{0}$. To ensure a unique characterization of $ \left\{\mathbf{P}_i(s,\boldsymbol{\theta_i})\right\}_{i=1}^K $, it is assumed that at most one submodel $ \mathbf{P}_i(s,\boldsymbol{\theta_i}) $ is biproper.  The $A_i(s)$ and $\mathbf{B}_i(s)$ polynomials are parametrized as
\begin{align} 
A_i(s) &= 1 + a_{i,1}s + \ldots + a_{i,n_i}s^{n_i}, \label{eq: FREQ - plant polynomial A}\\
\mathbf{B}_i(s) &= \mathbf{B}_{i,0} + \mathbf{B}_{i,1} s+ \ldots + \mathbf{B}_{i,m_i}s^{m_i}, \label{eq: FREQ - plant polynomial B}
\end{align}
where the $A_i(s)$ polynomials are stable, i.e., all roots lie in the left-half plane, and they do not share any common roots. The polynomials $A_i(s)$ and $\mathbf{B}_i(s)$ are jointly described by the parameter vector
\begin{align} \label{eq: FREQ - parameter vector}
\boldsymbol{\beta} = \begin{bmatrix} \boldsymbol{\theta}^{\top}_1 & \dots & \boldsymbol{\theta}^{\top}_K \end{bmatrix}^{\top},
\end{align}
where $\boldsymbol{\theta}_i$ for $i = 1,\ldots,K$ contains the parameters of the $i$th submodel
\begin{align} \label{eq: FREQ - parameter vector submodel}
\boldsymbol{\theta}_i = 
\begin{bmatrix}
a_{i,1} & \dots & a_{i,n_i} & 
\operatorname{vec}\left(\mathbf{B}_{i,0}\right)^{\top} & \dots & 
\operatorname{vec}(\mathbf{B}_{i,m_i})^{\top}
\end{bmatrix}^{\top}.
\end{align}
The identification problem is formulated based on the matrix residual, which is computed as the difference between the FRF dataset \( \{  \mathbf{G}(\omega_k) \}^N_{k=1}\) and the model, according to
\begin{equation} \label{eq: FREQ - residual}
\mathbf{E}\left(\omega_k, \boldsymbol{\beta}\right) = \mathbf{G}\left(\omega_k\right) - \mathbf{P}\left(\xi_k, \boldsymbol{\beta} \right),
\end{equation}
where $\xi_k = j\omega_k$. The parameter vector estimate $\hat{\boldsymbol{\beta}}$ is obtained as the minimizer of the weighted least-squares criterion
\begin{equation} \label{eq: FREQ - optimization problem}
\hat{\boldsymbol{\beta}}=\underset{\boldsymbol{\beta}}{\arg \min }\frac{1}{2N}\sum_{k=1}^N\left\| \operatorname{vec}\Bigl(\mathbf{E}\left(\omega_k, \boldsymbol{\beta}\right)\Bigr) \right\|_{\mathbf{W}(\omega_k)}^2,
\end{equation}
where $\mathbf{W}(\omega_k) \in \mathbb{C}^{n_{\mathrm{u}}n_{\mathrm{y}} \times n_{\mathrm{u}}n_{\mathrm{y}}}$ is a frequency-dependent weighting matrix. The main objective of the first stage is to estimate additive models, as described by \eqref{eq: FREQ - true plant}, that minimize the cost function in \eqref{eq: FREQ - optimization problem}, given the noisy dataset $\mathbf{G}(\omega_k)$.

\subsection{Stage 2: Structured identification of modal models}
The solution to the stage one identification problem yields an estimate of the additive parameter vector $ \hat{\boldsymbol{\beta}} $. Since the numerator matrices are unconstrained in the additive model structure, the resulting model generally cannot be represented exactly in modal form and has a higher order than necessary to describe the system \cite{Gustavsen2004ADomain}. To address this, the estimate \( \hat{\boldsymbol{\beta}} \) is projected onto the space of models that satisfy the rank constraint by solving a second least-squares problem. This procedure corresponds to the {Indirect Prediction Error Method (IPEM)}~\cite{Soderstrom1991AnIdentification}, which treats the intermediate estimate \( \hat{\boldsymbol{\beta}} \) as data. The IPEM framework enables structured model identification by exploiting the statistical information in \( \hat{\boldsymbol{\beta}} \) without requiring access to the original dataset. 

\begin{figure}[b]
        \centering
        \includegraphics[width=\linewidth]{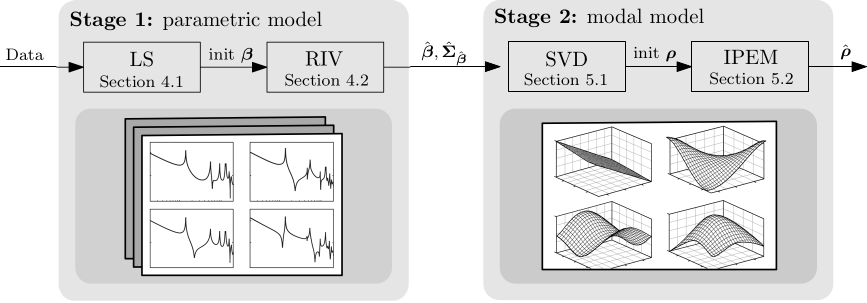}
        \caption{Outline of the estimation procedure. In Stage 1, an additive parametric model is estimated by fitting directly to the measured FRF data. A linear LS problem is first solved to obtain an initial estimate, which is subsequently refined by the RIV method via iterative minimization of a nonlinear cost function. In Stage 2, the modal model is obtained directly from the Stage 1 solution. The IPEM problem is initialized using SVDs, which subsequently projects the Stage 1 solution onto the set of models satisfying $\boldsymbol{\beta} \in \mathcal{S}_2$, yielding the final modal model estimate.}
        \label{fig: outline}
\end{figure}

Consider the nested model structures $ \mathcal{M}_1 $ and $ \mathcal{M}_2 $, where $ \mathcal{M}_1 $ denotes the modal model structure and $ \mathcal{M}_2 $ the additive model structure, with $ \mathcal{M}_1 \subset \mathcal{M}_2 $. Since the model structures $\mathcal{M}_1$ and $\mathcal{M}_2$ are nested, there exists a smooth and injective map $\mathbf{f}: \mathcal{S}_1 \to \mathcal{S}_2$, that links the parameter vectors $\boldsymbol{\rho}$ and $\boldsymbol{\beta}$ such that $\partial {\mathbf{f}}({\boldsymbol{\rho}}) / \partial {\boldsymbol{\rho}}^\top$ has full rank over $\mathcal{S}_1$. For the defined model structures, the subspace $\mathcal{S}_1 = \mathbb{R}^{\operatorname{dim}(\boldsymbol{\rho})}$ consists of the vectors $\boldsymbol{\rho}$ as defined in \eqref{eq: modal parameter vector}, while the subspace $\mathcal{S}_2 \subset \mathbb{R}^{\operatorname{dim}(\boldsymbol{\beta})}$ consists of the vectors $\boldsymbol{\beta}$ that admit a lower dimensional representation of the form $\mathbf{f}(\boldsymbol{\rho}) = \boldsymbol{\beta}$, i.e., the set of additive models satisfying the rank constraint. {The explicit form of $\mathbf{f}$ depends on the considered damping model and is further specified in Section~5.} An estimate of the parameter vector \( \boldsymbol{\rho} \) is then obtained as a solution to a weighted least-squares optimization problem
\begin{align} \label{eq: IPEM estimation problem}
\hat{\boldsymbol{\rho}} = \underset{\boldsymbol{\rho}}{\arg \min} 
\left\| \hat{\boldsymbol{\beta}} - \mathbf{f}(\boldsymbol{\rho}) \right\|_{\hat{\boldsymbol{\Sigma}}_{\hat{\boldsymbol{\beta}}}^{-1}}^2,
\end{align}
where \( \hat{\boldsymbol{\Sigma}}_{\hat{\boldsymbol{\beta}}} \) is an estimate of the parameter covariance matrix of the vector \( \hat{\boldsymbol{\beta}} \). Weighting by the parameter covariance ensures that the cost function in \eqref{eq: IPEM estimation problem} approximates the original cost function when expressed in terms of the modal parameter vector, thereby preserving the connection to the original dataset \cite{Soderstrom1991AnIdentification}. An expression for estimating \( \hat{\boldsymbol{\Sigma}}_{\hat{\boldsymbol{\beta}}} \) is provided in Section 4. The main objective of this second stage is to compute a solution \( \hat{\boldsymbol{\rho}} \) to \eqref{eq: IPEM estimation problem}, given \( \hat{\boldsymbol{\beta}} \) and its covariance estimate.
\begin{remark}
    To remove the ambiguity in mode-shape vector scaling, an additional constraint must be imposed on the mode-shape vectors in \eqref{eq: IPEM estimation problem}. If a collocated actuator-sensor pair is present in the system to be identified, the mass-normalized mode-shape vectors are uniquely determined \cite{Tacx2025Spatio-temporalControl}. However, in the absence of such a pair, as in the prototype wafer-stage system, this issue is resolved by either fixing one of the coefficients or including a norm constraint on one of the mode-shape vectors.
\end{remark}

\subsection{Discussion}
The proposed two-stage identification approach provides a computationally efficient framework for estimating structured modal models. By omitting the challenging rank constraint in the first stage, the initial estimation problem becomes considerably more tractable. The constraint is subsequently enforced through a second projection problem, which is easier to solve for two main reasons. First, the estimated additive model is often close to the modal subspace, enabling accurate initialization of the second-stage, which reduces the possibility of convergence to local minima. Second, the dimensionality of the data involved in the IPEM problem is significantly reduced. Instead of refitting the full dataset, the IPEM problem operates on the parameter vector \( \hat{\boldsymbol{\beta}} \) obtained from the first stage. Since \( \operatorname{dim}(\hat{\boldsymbol{\beta}}) \) is typically much smaller than the total number of frequency response data points \( n_{\mathrm{y}} n_{\mathrm{u}} N \), the optimization problem involves fewer data points. Consequently, the cost function tends to be better conditioned, making it easier to handle challenging model parameterizations and resulting in faster execution times.

\subsection{Outline of the estimation procedure}
In the following sections, the solution procedures are provided for the identification problems \eqref{eq: FREQ - optimization problem} and \eqref{eq: IPEM estimation problem}. Figure \ref{fig: outline} provides a schematic overview of the two-stage estimation procedure, indicating the subsections in which the solution methods for each stage are discussed. In the first stage, a linear least squares (LS) problem is solved to obtain an initial estimate, which is subsequently used as starting point for a Refined Instrumental Variable (RIV) method. In the second stage, the additive model is used to derive an initial modal model via Singular Value Decompositions (SVDs), yielding a suboptimal estimate that is then refined by solving the IPEM problem to obtain the final modal model estimate.

\section{Frequency-domain identification of MIMO additive models: A refined instrumental variable approach}
This section introduces a RIV method for estimating MIMO additive models and thereby provides contribution C1. First, an initialization procedure for generating initial estimates is presented, followed by the RIV estimation algorithm.
\subsection{Initialization via linear least-squares}
To solve the problem \eqref{eq: FREQ - optimization problem} an initial estimate of the model parameters $\boldsymbol{\beta}$ is needed. A convex linear least-squares method is introduced, which allows the numerator parameters to be computed in closed form, assuming fixed denominator polynomials. This reduces the initialization problem to determining initial pole locations, which are often effectively obtained from finite element models, FRF data, or tools such as complex mode indicator functions (CMIF) \cite{Shih1988ComplexEstimation}. To this end, assume that the denominator polynomials are fixed at $\bar{A}_i(s)$, and let $\boldsymbol{\eta}$ represent the parameter vector from \eqref{eq: FREQ - parameter vector} without the denominator parameters. The estimate $\hat{\boldsymbol{\eta}}$ is found as the solution to the convex problem
\begin{equation} \label{eq: convex problem}
    \hat{\boldsymbol{\eta}} = \underset{\boldsymbol{\eta}}{\arg \min }\frac{1}{2N}\sum^N_{k=1}\bigl\|\operatorname{vec}\bigl(\mathbf{G}(\omega_k)\bigr)-\boldsymbol{\Phi}^{\top}\left(\omega_k\right) \boldsymbol{\eta}\bigr\|_{\mathbf{W}(\omega_k)}^2,
\end{equation}
with the regressor matrix $\boldsymbol{\Phi}$ given by
\begin{equation} \label{eq: FREQ regressor}
\mathbf{\Phi}\left(\omega_k\right) = 
\left[\begin{array}{ccc}
\mathbf{\Phi}_{1}^\top\left(\omega_k\right) & \ldots & \mathbf{\Phi}_{K}^\top\left(\omega_k\right)
\end{array}\right]^\top, \quad
\mathbf{\Phi}_{i}\left(\omega_k\right) = 
\begin{bmatrix}
\displaystyle\frac{\mathbf{I}_{n_{\mathrm{u}}n_{\mathrm{y}}}}{\xi^{\ell_i} \bar{A}_i(\xi_k)} &
\ldots &
\displaystyle\frac{\xi^{m_i} \mathbf{I}_{n_{\mathrm{u}}n_{\mathrm{y}}}}{\xi^{\ell_i} \bar{A}_i(\xi_k)}
\end{bmatrix}^\top.
\end{equation}
Hence, an initial estimate of ${\boldsymbol{\beta}}$ is determined by first providing initial pole locations, which enables the computation of the numerator parameters as the solution to the convex problem \eqref{eq: convex problem}.

\subsection{RIV estimation for additive systems}
Next, the RIV estimation method used to solve the nonconvex optimization problem in~\eqref{eq: FREQ - optimization problem} is introduced. First, the optimality conditions used in the method are established. The estimator is subsequently derived and finally an expression for the parameter covariance is provided.

\subsubsection{Criterion for optimality}
The minimizers of the cost function in \eqref{eq: FREQ - optimization problem} satisfy the first-order optimality condition 
\begin{equation} \label{eq: FREQ - first-order optimality}
\begin{aligned}
\mathbf{0} = \frac{1}{N}\sum_{k=1}^N \Re \left\{ \hat{\mathbf{\Phi}}\left(\omega_k, \boldsymbol{\beta}\right) \mathbf{W}(\omega_k) \operatorname{vec}\bigl(\mathbf{E}\left(\omega_k, \boldsymbol{\beta}\right)\bigr) \right\},
\end{aligned}
\end{equation}
with the Jacobian matrix
\begin{equation}
\hat{\mathbf{\Phi}}\left(\omega_k, \boldsymbol{\beta}\right) = \left(\frac{\partial \operatorname{vec}\bigl(\mathbf{E}\left(\omega_k, \boldsymbol{\beta}\right)\bigr)}{\partial \boldsymbol{\beta}^\top}\right)^\hop.     
\end{equation}
For the considered additive model parameterization the Jacobian corresponds to
\begin{equation} \label{eq: FREQ - instrument matrix}
    \hat{\mathbf{\Phi}}\left(\omega_k, \boldsymbol{\beta}\right) = \Bigl[ \begin{array}{ccc} \hat{\mathbf{\Phi}}_{1}^\hop\left(\omega_k, \boldsymbol{\theta}_1\right) & \ldots & \hat{\mathbf{\Phi}}_{K}^\hop\left(\omega_k, \boldsymbol{\theta}_K\right)\end{array}\Bigr]^\hop, 
\end{equation}
where $\hat{\mathbf{\Phi}}_{i}\left(\omega_k, \boldsymbol{\theta}_i\right)$ for $i = 1,\ldots,K$ is given by
\begin{equation} \label{eq: FREQ - instrument matrix submodel}
\hat{\boldsymbol{\Phi}}_{i}(\omega_k,\boldsymbol{\beta}) = 
\begin{bmatrix}
\displaystyle\frac{-\xi_k \operatorname{vec}(\mathbf{P}_i(\xi_k, \boldsymbol{\theta}_i))}{\xi_k^{\ell_i} A_i(\xi_k)} &
\ldots &
\displaystyle\frac{-\xi_k^{n_i} \operatorname{vec}(\mathbf{P}_i(\xi_k, \boldsymbol{\theta}_i))}{\xi_k^{\ell_i} A_i(\xi_k)} &
\displaystyle\frac{\mathbf{I}_{n_{\mathrm{u}}n_{\mathrm{y}}}}{\xi_k^{\ell_i} A_i(\xi_k)} &
\ldots &
\displaystyle\frac{\xi_k^{m_i} \mathbf{I}_{n_{\mathrm{u}}n_{\mathrm{y}}}}{\xi_k^{\ell_i} A_i(\xi_k)}
\end{bmatrix}^{\hop}.
\end{equation}
Next, the first-order optimality condition \eqref{eq: FREQ - first-order optimality} will be exploited to derive an estimator for the parameter vector $\boldsymbol{\beta}$.

\subsubsection{Derivation of the RIV estimator}
The condition in (\ref{eq: FREQ - first-order optimality}) is non-linear in the parameter vector $\boldsymbol{\beta}$. A solution is obtained by reformulating \eqref{eq: FREQ - residual} to a pseudolinear form which enables the RIV approach. For each submodel in the additive model structure, the residual can be reformulated into an unique pseudolinear regression, as stated in the following lemma. 
\begin{lemma} The pseudolinear regression form of the residual \eqref{eq: FREQ - residual} corresponding to the $i$th submodel is expressed as
\begin{equation} \label{eq: FREQ - subproblems regression form}
\operatorname{vec}\bigl(\mathbf{E}\left(\omega_k, \boldsymbol{\beta}\right)\bigr)  =\operatorname{vec}(\tilde{\mathbf{G}}_{f,i}(\omega_k, \boldsymbol{\beta})) -\boldsymbol{\Phi}_{i}^{\top}\left(\omega_k, \boldsymbol{\beta}\right) \boldsymbol{\theta}_i, 
\end{equation}
with the regressor
\begin{equation}
\boldsymbol{\Phi}_i(\omega_k,\boldsymbol{\beta}) = 
\begin{bmatrix}
\displaystyle\frac{-\xi_k\operatorname{vec}(\tilde{\mathbf{G}}_i(\omega_k, \boldsymbol{\beta}))}{A_i(\xi_k)} &
\ldots &
\displaystyle\frac{-\xi_k^{n_i}\operatorname{vec}(\tilde{\mathbf{G}}_i(\omega_k, \boldsymbol{\beta}))}{A_i(\xi_k)} &
\displaystyle\frac{\mathbf{I}_{n_{\mathrm{u}}n_{\mathrm{y}}}}{\xi_k^{\ell_i} A_i(\xi_k)} &
\ldots &
\displaystyle\frac{\xi_k^{m_i} \mathbf{I}_{n_{\mathrm{u}}n_{\mathrm{y}}}}{\xi_k^{\ell_i} A_i(\xi_k)}
\end{bmatrix}^\top,
\end{equation}
and where $\operatorname{vec}(\tilde{\mathbf{G}}_{f,i}(\omega_k, \boldsymbol{\beta})) = A^{-1}_i(\xi_k)\operatorname{vec}(\tilde{\mathbf{G}}_i(\omega_k, \boldsymbol{\beta}))$ with $\operatorname{vec}(\tilde{\mathbf{G}}_i(\omega_k, \boldsymbol{\beta}))$ the  residual plant of the $i$th submodel, defined by
\begin{equation} \label{eq: FREQ - residual plant}
\tilde{\mathbf{G}}^{}_{i}\left(\omega_k, \boldsymbol{\beta}\right) = \mathbf{G}^{}(\omega_k) - \sum_{\substack{\ell=1, \ldots, K \\ \ell \neq i}} \mathbf{P}_\ell(\xi_k, \boldsymbol{\theta}_\ell).
\end{equation}
\end{lemma}
\begin{proof} \textit{
The residual \eqref{eq: FREQ - residual} is rewritten for $i = 1,\ldots,K$ according to
\begin{align}
\mathbf{E}\left(\omega_k, \boldsymbol{\beta} \right) & =\tilde{\mathbf{G}}_i\left(\omega_k,\boldsymbol{\beta}\right)-\frac{\mathbf{B}_i\left(\xi_k\right)}{\xi_k^{\ell_i}A_i\left(\xi_k\right)}, \\
& =\frac{1}{\xi_k^{\ell_i}A_i\left(\xi_k\right)}\left(\xi_k^{\ell_i}A_i\left(\xi_k\right) \tilde{\mathbf{G}}_i\left(\omega_k,\boldsymbol{\beta}\right)-\mathbf{B}_i\left(\xi_k\right)\right), \label{eq: FREQ - subproblems}
\end{align}
with $\tilde{\mathbf{G}}_i$ defined in \eqref{eq: FREQ - residual plant}. Substituting the numerator and denominator polynomials \eqref{eq: FREQ - plant polynomial A} and \eqref{eq: FREQ - plant polynomial B}, and vectorizing both sides, \eqref{eq: FREQ - subproblems} yields
\begin{align}
\operatorname{vec}\bigl(\mathbf{E}\left(\omega_k, \boldsymbol{\beta}\right)\bigr) &=  \frac{\operatorname{vec}(\tilde{\mathbf{G}}_{i}(\omega_k, \boldsymbol{\beta}))}{A_i\left(\xi_k\right)} + \ldots + \frac{a_{n_i}\xi_k^{n_i} \operatorname{vec}(\tilde{\mathbf{G}}_{i}(\omega_k, \boldsymbol{\beta}))}{A_i\left(\xi_k\right)} \notag  -\frac{\operatorname{vec}\left(\mathbf{B}_{i, 0}\right)}{\xi_k^{\ell_i}A_i\left(\xi_k\right)}-\ldots-\frac{\xi_k^{m_i} \operatorname{vec}\left(\mathbf{B}_{i, m}\right)}{\xi_k^{\ell_i}A_i\left(\xi_k\right)}.
\end{align}
This expression can directly be written in the form \eqref{eq: FREQ - subproblems regression form} using \eqref{eq: FREQ - parameter vector submodel}, thereby completing the proof.}
\end{proof}
The residual formulation in \eqref{eq: FREQ - subproblems regression form} defines $K$ pseudolinear regressions. Introducing the stacked signals
\begin{align}
    \mathbf{\Upsilon}\left(\omega_k, \boldsymbol{\beta}\right) &= \Bigl[ \begin{array}{ccc}  \operatorname{vec}(\tilde{\mathbf{G}}_{f,1}(\omega_k, \boldsymbol{\beta})) & \dots & \operatorname{vec}(\tilde{\mathbf{G}}_{f,K}(\omega_k, \boldsymbol{\beta}))\end{array}\Bigl]^\top, \label{eq: FREQ - stacked residual output} \\ 
    \mathbf{\Phi}\left(\omega_k, \boldsymbol{\beta}\right) &= \Bigl[ \begin{array}{ccc} \mathbf{\Phi}^\top_{1}\left(\omega_k, \boldsymbol{\beta}\right) & \dots & \mathbf{\Phi}^\top_{K}\left(\omega_k, \boldsymbol{\beta}\right)\end{array}\Bigr]^\top \label{eq: FREQ - stacked regressor matrix},
\end{align}
and the parameter matrix
\begin{equation} \label{eq: FREQ - parameter matrix}
\mathcal{B}=\left[\begin{array}{ccc}
\boldsymbol{\theta}_1 & & \mathbf{0} \\
& \ddots & \\
\mathbf{0} & & \boldsymbol{\theta}_K
\end{array}\right],
\end{equation}
which contains the elements of $\boldsymbol{\beta}$ along the block diagonal, allows to write the equivalent optimality condition \eqref{eq: FREQ - first-order optimality} for the $K$ subproblems as
\begin{equation} \label{eq: FREQ - instrumental variable equation - matrix}
\begin{aligned}
    \sum_{k=1}^N \Re \left\{\hat{\mathbf{\Phi}}(\omega_k, \boldsymbol{\beta})\mathbf{W}(\omega_k) \Bigl( \mathbf{\Upsilon}^\top(\omega_k, \boldsymbol{\beta}) - \mathbf{\Phi}^\top(\omega_k,\boldsymbol{\beta})\mathcal{B} \Bigr) \right\} = \mathbf{0}.
\end{aligned}
\end{equation}
The solution to \eqref{eq: FREQ - instrumental variable equation - matrix} is found iteratively by fixing $\boldsymbol{\beta} = \boldsymbol{\beta}^{\langle j \rangle}$ at the $j$th iteration in (\ref{eq: FREQ - stacked residual output}), the regressor (\ref{eq: FREQ - stacked regressor matrix}), and additionally the Jacobian (\ref{eq: FREQ - instrument matrix}), which leads to the following iterative procedure.
\begin{algorithm2}
Given an initial estimate $\boldsymbol{\beta}^{\langle 0\rangle}$ and tolerance $\epsilon_1$, compute a new estimate until
$
\|\boldsymbol{\beta}^{\langle j+1\rangle} - \boldsymbol{\beta}^{\langle j\rangle}\| / \|\boldsymbol{\beta}^{\langle j\rangle}\| < \epsilon_1
$
using:
\begin{align} \label{eq: FREQ - estimator}
\mathcal{B}^{\langle j+1 \rangle} = {\left[ \sum_{k=1}^N \hat{\boldsymbol{\Phi}}(\omega_k, \boldsymbol{\beta}^{\langle j \rangle}) \mathbf{W}(\omega_k) \boldsymbol{\Phi}^{\top}(\omega_k, \boldsymbol{\beta}^{\langle j \rangle})\right]^{-1} } \sum_{k=1}^N \hat{\boldsymbol{\Phi}}(\omega_k, \boldsymbol{\beta}^{\langle j \rangle}) \mathbf{W}(\omega_k) \boldsymbol{\Upsilon}^{\top}(\omega_k, \boldsymbol{\beta}^{\langle j \rangle}),
\end{align}
where the updated parameter vector $\boldsymbol{\beta}^{\langle j+1 \rangle}$ is extracted from the block-diagonal coefficients of $\mathcal{B}^{\langle j+1 \rangle}$, as described in~(\ref{eq: FREQ - parameter matrix}).
\end{algorithm2}
The convergence point of the iterations described by (\ref{eq: FREQ - estimator}) satisfies the first-order optimality condition in (\ref{eq: FREQ - first-order optimality}). As a result, the estimate corresponds to a stationary point of the cost function defined in (\ref{eq: FREQ - optimization problem}), thereby ensuring local optimality. Since the original cost function is nonconvex, accurate initialization is critical to ensure convergence to the global optimum.
\begin{remark}
As stability is not explicitly enforced in the estimator \eqref{eq: FREQ - estimator}, the resulting model may contain unstable poles. A common approach for RIV estimators to address this is to reflect any unstable continuous-time poles across the imaginary axis at each iteration \citep{Garnier2008IdentificationData}. Alternatively, if each submodel has a denominator of order $n_i \leq 2$, stability is directly enforced by constraining the denominator coefficients to be positive.
\end{remark}
\begin{remark} Note that the iterations described by \eqref{eq: FREQ - estimator} correspond to a RIV method, where $\hat{\mathbf{\Phi}}$ is interpreted as the instrument matrix \citep{Young1980RefinedAnalysis}. Furthermore, for $K=1$, the iterations in (\ref{eq: FREQ - estimator}) correspond to the frequency-domain RIV method in \cite{Blom2010MultivariableRegression}, and by replacing $\hat{\mathbf{\Phi}}$ with $\mathbf{\Phi}$ to the SK iterations by \cite{Sanathanan1961TransferPolynomials}. The method can be considered as a MIMO frequency-domain variant of the SISO approach introduced in \cite{Gonzalez2024IdentificationClosed-loop}.
\end{remark}

\subsubsection{Estimator for the parameter covariance}
The uncertainty of the estimates is quantified using the parameter covariance matrix. The expression for estimating this covariance matrix is derived following a similar approach to that presented in \cite[Chapter 9]{Ljung1999SystemUser} and \cite{Gonzalez2025StatisticallyIdentification}; a detailed derivation of this result is beyond the scope of this paper. Specifically, an approximation of the parameter covariance of the estimator in \eqref{eq: FREQ - estimator} is given by  
\begin{equation} \label{eq: cov}
\hat{\boldsymbol{\Sigma}}_{\hat{\boldsymbol{\beta}}}
=
\left[
\frac{1}{N}
\sum_{k=1}^{N}
\hat{\boldsymbol{\Phi}}(\omega_{k},\,\hat{\boldsymbol{\beta}})
\;\hat{\boldsymbol{\Sigma}}_{\hat{\mathbf{G}}}(\omega_{k})
\;\hat{\boldsymbol{\Phi}}^{\hop}(\omega_{k},\,\hat{\boldsymbol{\beta}})
\right]^{-1}.
\end{equation}
where $\hat{\boldsymbol{\beta}}$ denotes the converged solution of \eqref{eq: FREQ - estimator}, $\hat{\boldsymbol{\Phi}}$ represents the Jacobian of \eqref{eq: FREQ - instrument matrix} evaluated at $\hat{\boldsymbol{\beta}}$, and $\hat{{\boldsymbol{\Sigma}}}_{\hat{\mathbf{G}}} (\omega_k)$ is an estimate of the covariance matrix of the measured FRF \cite{Pintelon2012SystemIdentification}. The estimate obtained using \eqref{eq: cov} serves as the weighting matrix in the IPEM problem specified in \eqref{eq: IPEM estimation problem}.

\section{Structured identification of modal models}
This section presents the solution procedure for the IPEM problem in~\eqref{eq: IPEM estimation problem}, thereby providing contribution C2. First, the approach for obtaining an initial modal model is discussed. Subsequently, a solution method for the structured identification problem is provided for both general-viscously damped and proportionally damped mechanical systems. The procedure outlined below assumes that an estimate \( \hat{\boldsymbol{\beta}} \), along with its corresponding covariance estimate $\boldsymbol{\hat{\Sigma}}_{\boldsymbol{\hat{\beta}}}$, is available. 

\subsection{Initial modal model via SVD decomposition}
To solve the IPEM problem, the estimated additive model is first used to derive an initial estimate of the modal parameter vector in \eqref{eq: modal parameter vector}. The eigenvalues can be computed exactly from the estimated denominator coefficients. The corresponding mode-shape vectors are initialized by reducing each estimated residue matrix to its best rank-one approximation. This is achieved via SVDs, which provides the optimal approximation in least-squares sense, as formalized below.
\begin{theorem}[Eckart--Young--Mirsky {\cite{Horn2013MatrixEdition}}]
Let \( \mathbf{A} \in \mathbb{R}^{n \times m} \) have SVD \( \mathbf{A} = \sum_{i=1}^r \sigma_i \mathbf{u}_i \mathbf{v}_i^\top \), with \( \sigma_1 \geq \cdots \geq \sigma_r > 0 \). The best rank-one approximation of \( \mathbf{A} \) in the Frobenius norm is \( \mathbf{A}_1 = \sigma_1 \mathbf{u}_1 \mathbf{v}_1^\top \).
\end{theorem}
\noindent This decomposition is applied to the residue matrices estimated in the first stage, where the parameterization ~\eqref{eq: modal transfer second order} is considered for the general case, and ~\eqref{eq: modal transfer second order propportional} for the proportionally damped case. The estimates of the real residue matrices \( \hat{\mathbf{R}}^{\mathrm{flex}}_i \) are directly available, whereas the complex residue matrices \( \hat{\mathbf{L}}_i \) are obtained by rearranging ~\eqref{eq: term 1} and \eqref{eq: term 2}, yielding  
\begin{equation}
    \hat{\mathbf{L}}_i = \frac{1}{\lambda_i - \bar{\lambda}_i} \left( \hat{\mathbf{N}}_{i,0} + \lambda_i \hat{\mathbf{N}}_{i,1} \right).
\end{equation}
The mode-shape vectors \( \boldsymbol{\psi}_{l,i} \) and \( \boldsymbol{\psi}_{r,i} \) are initialized using the dominant left and right singular vectors of the SVD decomposition \( \hat{\mathbf{L}}_i = \mathbf{U} \boldsymbol{\Sigma} \mathbf{V}^\hop \). In the additive model structure, all rigid-body modes are represented collectively by a single submodel, whereas in the modal representation they appear as a sum over \( n_{\mathrm{rbm}} \) individual rigid-body modes. As a result, the corresponding residue matrix \( \hat{\mathbf{R}}^{\mathrm{rbm}} \approx \sum_{j=1}^{n_{\mathrm{rbm}}}\mathbf{R}^{\mathrm{rbm}}_j \) contains the combined contribution of all rigid-body modes, with its dominant singular directions spanning the rigid-body subspace. Consequently, the initial estimates for the rigid-body mode-shape vectors \( \boldsymbol{\phi}_{l,j} \) and \( \boldsymbol{\phi}_{r,j} \) are obtained from the first \( n_{\mathrm{rbm}} \) dominant singular vectors of the SVD \( \hat{\mathbf{R}}^{\mathrm{rbm}} = \mathbf{U}\boldsymbol{\Sigma}\mathbf{V}^\hop \). The mode-shape vectors for the proportional damping case are initialized following the same procedure.

\subsection{IPEM for modal model estimation}
{A key aspect of the IPEM problem~\eqref{eq: IPEM estimation problem} is the definition of the function $\mathbf{f}: \boldsymbol{\rho} \mapsto \boldsymbol{\beta}$ which links the modal parameter vector~\eqref{eq: modal parameter vector} to the additive parameter vector~\eqref{eq: FREQ - parameter vector}. The function \( \mathbf{f} \) specifies how a given set of modal parameters is translated into the corresponding coefficients of the additive model representation. In essence, \( \boldsymbol{\rho} \) and \(\mathbf{f}(\boldsymbol{\rho}) = \boldsymbol{\beta} \) describe the same system but expressed in different parameterizations, and \( \mathbf{f} \) provides the formal relationship between these two representations.} For the considered model structures, the mapping \( \mathbf{f} \) allows the partitioning
\begin{align} \label{eq: parameter map}
\mathbf{f}(\boldsymbol{\rho}) = 
\begin{bmatrix}
\mathbf{g}(\boldsymbol{\varrho}_1, \dots, \boldsymbol{\varrho}_{n_\mathrm{rbm}}) &
\mathbf{h}_1(\boldsymbol{\rho}_1) &
\ldots &
\mathbf{h}_{n_{\mathrm{flex}}}(\boldsymbol{\rho}_{n_{\mathrm{flex}}})
\end{bmatrix}^{\top},
\end{align}
where the function \( \mathbf{g}: [\boldsymbol{\varrho}_1^\top, \ldots, \boldsymbol{\varrho}_{n_{\mathrm{rbm}}}^\top]^\top \mapsto \boldsymbol{\theta}_1 \) maps the rigid-body mode parameter vectors of the modal representation to the corresponding additive model parameter vector, and each \( \mathbf{h}_i: \boldsymbol{\rho}_i \mapsto \boldsymbol{\theta}_i \) maps the parameter vector describing the \( i \)th flexible mode from modal representation to additive model representation. The explicit definition of the parameter mappings depend on the considered damping model and are provided in the subsequent subsections. A solution to the nonconvex problem in~\eqref{eq: IPEM estimation problem} can be obtained, e.g., using a Gauss--Newton (GN) procedure. At each iteration \( j \), the residual  
\begin{equation}
\boldsymbol{\varepsilon}(\boldsymbol{\rho}) = \hat{\boldsymbol{\beta}} - \mathbf{f}(\boldsymbol{\rho}),
\end{equation}
is linearized by a first-order Taylor expansion. This yields a local quadratic approximation of the cost function in terms of the parameter perturbation \( \Delta_{\boldsymbol{\rho}} \), from which the optimal update is computed via linear regression. Thus, the GN iterations proceed as follows.
\begin{algorithm2}  Given an initial estimate $\boldsymbol{\rho}^{\langle 0\rangle}$ and tolerance $\epsilon_2$, compute a new estimate until $\|\boldsymbol{\rho}^{\langle j+1\rangle} - \boldsymbol{\rho}^{\langle j\rangle}\| / \| \boldsymbol{\rho}^{\langle j\rangle} \| < \epsilon_2 $ by solving 
\begin{align} \label{eq: GN iterations}
\hat{\boldsymbol{\rho}}^{\langle j+1\rangle} = \hat{\boldsymbol{\rho}}^{\langle j\rangle} + \alpha \underset{\Delta_{\boldsymbol{\rho}}}{\arg \min} \left\| \boldsymbol{\varepsilon}(\boldsymbol{\rho}^{\langle j\rangle}) - \mathbf{J}(\boldsymbol{\rho}^{\langle j\rangle}) \Delta_{\boldsymbol{\rho}} \right\|_{\hat{\boldsymbol{\Sigma}}^{-1}_{\hat{\boldsymbol{\beta}}}}^2, 
\end{align}
where $\alpha$ is a step-size parameter and {$\mathbf{J}(\boldsymbol{\rho}^{\langle j\rangle})$ is the analytic Jacobian matrix evaluated at the current parameter estimate}
\begin{equation} \label{eq: jacobian}
\mathbf{J}(\boldsymbol{\rho}^{\langle j\rangle}) = \left.\frac{\partial \mathbf{f}\left(\boldsymbol{\rho}\right)}{\partial \boldsymbol{\rho}^{\top}}\right|_{\hat{\boldsymbol{\rho}}^{\langle j\rangle}}. 
\end{equation}
\end{algorithm2}
\noindent The converged solution of the iterations in~\eqref{eq: GN iterations} provides a locally optimal estimate of the nonlinear least-squares problem~\eqref{eq: IPEM estimation problem}. {Since the cost function is nonconvex, global optimality cannot be guaranteed and the obtained solution depends on the chosen initialization.} The following subsections provide the explicit forms of the parameter mappings, along with their associated Jacobians, for each of the considered model structures.

\subsubsection{Parameter mappings for general-viscously damped mechanical systems}
{
For the general damping model, the mappings \( \mathbf{g} \) and \( \mathbf{h}_i \) are derived considering the second-order parameterization in~\eqref{eq: modal transfer second order}, with the associated parameter vector defined in~\eqref{eq: modal parameter vector}. The function \( \mathbf{g} \) links the modal and additive parameterizations of the rigid-body modes, and since the additive representation describes all rigid-body dynamics using a single submodel, \( \mathbf{g} \) amounts to summing the dyads of the \( n_{\mathrm{rbm}} \) rigid-body mode-shape vectors according to
\begin{equation} \label{eq: parameter map 1}
\mathbf{g}(\boldsymbol{\varrho}_{1},\dots,  \boldsymbol{\varrho}_{n_\mathrm{rbm}}) = \operatorname{vec}\left( \sum_{j=1}^{n_{\mathrm{rbm}}} \boldsymbol{\phi}^{}_{l,j}  \boldsymbol{\phi}^\top_{r,j} \right).
\end{equation}  
The function \( \mathbf{h}_i \) links the modal and additive parameterizations of the \( i \)th flexible mode, with \( i = 1,\ldots,n_{\mathrm{flex}} \). The mapping includes the relation between eigenvalues and quadratic denominator parameters and the construction of the rank-2 numerator matrices from the complex mode-shape vectors. Specifically,
\begin{align} \label{eq: parameter map 2}
\mathbf{h}_{i}(\boldsymbol{\rho}_{i}) = 
\begin{bmatrix}
\lambda_i \bar{\lambda}_i &
-\left(\lambda_i + \bar{\lambda}_i\right) &
-\operatorname{vec}\left( \bar{\lambda}_i \boldsymbol{\psi}_{l,i} \boldsymbol{\psi}^\top_{r,i} + \lambda_i \bar{\boldsymbol{\psi}}_{l,i} \bar{\boldsymbol{\psi}}^\top_{r,i} \right)^\top &
\operatorname{vec}\left( \boldsymbol{\psi}_{l,i} \boldsymbol{\psi}^\top_{r,i} + \bar{\boldsymbol{\psi}}_{l,i} \bar{\boldsymbol{\psi}}^\top_{r,i} \right)^\top
\end{bmatrix}^{\top},
\end{align}
where the first two entries correspond to the denominator coefficients, and the last two entries correspond to the numerator matrices.}   Since each mode is parameterized independently, the Jacobian in~\eqref{eq: jacobian} takes a block-diagonal form
\begin{align}
\mathbf{J}(\boldsymbol{\rho}) = 
\begin{bmatrix}
\mathbf{S}(\boldsymbol{\varrho}_{1},\dots,  \boldsymbol{\varrho}_{n_\mathrm{rbm}}) & \mathbf{0} & \cdots & \mathbf{0} \\
\mathbf{0} & \mathbf{X}_1(\boldsymbol{\rho}_1) & \cdots & \mathbf{0}  \\
\vdots & \vdots & \ddots & \vdots  \\
\mathbf{0} & \mathbf{0} & \cdots & \mathbf{X}_{n_{\mathrm{flex}}}(\boldsymbol{\rho}_{n_{\mathrm{flex}}})  \\
\end{bmatrix},
\end{align}
where \( \mathbf{S} \) denotes the Jacobian of~\eqref{eq: parameter map 1}, and \( \mathbf{X}_i \) corresponds to the Jacobian of~\eqref{eq: parameter map 2}. The derivation of the Jacobians relies on the following identity, applicable to vectors and matrices of compatible dimensions
\begin{align*}
\operatorname{vec}(\mathbf{u} \mathbf{x}^{\top}) & = \left(\mathbf{x} \otimes \mathbf{I}\right)\mathbf{u} = \left(\mathbf{I} \otimes \mathbf{u} \right)\mathbf{x},
\end{align*}
which allows dyadic vector products to be expressed in an affine form. Working out the derivatives leads for \( \mathbf{S} \) to 
\begin{align}
\mathbf{S}(\boldsymbol{\varrho}_1, \dots, \boldsymbol{\varrho}_{n_\mathrm{rbm}}) = 
\begin{bmatrix}
\boldsymbol{\phi}_{r,1} \otimes \mathbf{I}_{n_{\mathrm{y}}} &
\mathbf{I}_{n_{\mathrm{u}}} \otimes \boldsymbol{\phi}_{l,1} &
\ldots &
\boldsymbol{\phi}_{r,n_{\mathrm{rbm}}} \otimes \mathbf{I}_{n_{\mathrm{y}}} &
\mathbf{I}_{n_{\mathrm{u}}} \otimes \boldsymbol{\phi}_{l,n_{\mathrm{rbm}}}
\end{bmatrix},
\end{align}
while the Jacobian \( \mathbf{X}_i \) pertaining to the \( i \)th flexible mode results in
\begin{equation}
\mathbf{X}_i(\boldsymbol{\rho}_i) = \begin{bmatrix}
2 \Re(\lambda_i) & 2 \Im(\lambda_i) & 0 & 0 & 0 & 0 \\
-2 & 0  & 0 & 0 & 0 & 0 \\ 
2\operatorname{vec}( \Re( \boldsymbol{\psi}_{l,i} \boldsymbol{\psi}_{r,i}^\top ) ) & 
-2\operatorname{vec}( \Im( \boldsymbol{\psi}_{l,i} \boldsymbol{\psi}_{r,i}^\top)) & 
\mathbf{X}_{i,33} & \mathbf{X}_{i,34} & \mathbf{X}_{i,35} & \mathbf{X}_{i,36} \\
0 & 0 & \mathbf{X}_{i,43} & \mathbf{X}_{i,44} & \mathbf{X}_{i,45} & \mathbf{X}_{i,46}
\end{bmatrix},
\end{equation}
with the submatrices \( \mathbf{X}_{i,jk} \) defined as
\[
\begin{array}{ll}
\begin{aligned}
\mathbf{X}_{i,33} &= -\bar{\lambda}_i (\boldsymbol{\psi}_{r,i} \otimes \mathbf{I}_{n_{\mathrm{y}}} ) - \lambda_i (\bar{\boldsymbol{\psi}}_{r,i} \otimes \mathbf{I}_{n_{\mathrm{y}}} ), \\
\mathbf{X}_{i,34} &= -j\bar{\lambda}_i  (\boldsymbol{\psi}_{r,i} \otimes \mathbf{I}_{n_{\mathrm{y}}} ) + j\lambda_i (\bar{\boldsymbol{\psi}}_{r,i}  \otimes \mathbf{I}_{n_{\mathrm{y}}} ), \\
\mathbf{X}_{i,43} &= (\boldsymbol{\psi}_{r,i} \otimes \mathbf{I}_{n_{\mathrm{y}}} ) + (\bar{\boldsymbol{\psi}}_{r,i}  \otimes \mathbf{I}_{n_{\mathrm{y}}} ), \\
\mathbf{X}_{i,44} &= j (\boldsymbol{\psi}_{r,i} \otimes \mathbf{I}_{n_{\mathrm{y}}} ) - j (\bar{\boldsymbol{\psi}}_{r,i}  \otimes \mathbf{I}_{n_{\mathrm{y}}} ),
\end{aligned}
&
\begin{aligned}
\mathbf{X}_{i,35} &= -\bar{\lambda}_i (\mathbf{I}_{n_{\mathrm{u}}} \otimes \boldsymbol{\psi}_{l,i}) - \lambda_i (\mathbf{I}_{n_{\mathrm{u}}} \otimes \bar{\boldsymbol{\psi}}_{l,i}), \\
\mathbf{X}_{i,36} &= -j\bar{\lambda}_i  (\mathbf{I}_{n_{\mathrm{u}}} \otimes \boldsymbol{\psi}_{l,i}) + j\lambda_i  (\mathbf{I}_{n_{\mathrm{u}}} \otimes \bar{\boldsymbol{\psi}}_{l,i}), \\
\mathbf{X}_{i,45} &= (\mathbf{I}_{n_{\mathrm{u}}} \otimes \boldsymbol{\psi}_{l,i}) + (\mathbf{I}_{n_{\mathrm{u}}} \otimes \bar{\boldsymbol{\psi}}_{l,i}),  \\
\mathbf{X}_{i,46} &= j (\mathbf{I}_{n_{\mathrm{u}}} \otimes \boldsymbol{\psi}_{l,i}) - j (\mathbf{I}_{n_{\mathrm{u}}} \otimes \bar{\boldsymbol{\psi}}_{l,i}).
\end{aligned}
\end{array}
\]
\subsubsection{Parameter mappings for proportionally damped mechanical systems}
For proportionally damped systems, the mapping functions \( \mathbf{g} \) and \( \mathbf{h}_i \) are derived considering the second-order parameterization in~\eqref{eq: modal transfer second order}, with the associated parameter vector defined in~\eqref{eq: modal parameter vector}. The rigid-body mapping \( \mathbf{g} \) is identical to the expression in~\eqref{eq: parameter map 1}. For the \( i \)th flexible mode, \( i = 1,\ldots,n_{\mathrm{flex}} \), the function \( \mathbf{h}_i \) takes the form
\begin{equation} \mathbf{h}_i\bigl(\boldsymbol{\rho}_i\bigr) = \begin{bmatrix} \omega_i^2 & 2\,\zeta_i\,\omega_i & \displaystyle \operatorname{vec}\bigl(\boldsymbol{\phi}_{l,i}\,\boldsymbol{\phi}_{r,i}^{\top}\bigr)^{\top} \end{bmatrix}, \end{equation}
where the first two entries define the denominator coefficients, while the final entry contains the numerator terms constructed from the real mode-shape vectors. Finally, the Jacobian matrix takes the same form as in \eqref{eq: jacobian}, but with the Jacobian $\mathbf{X}_i$ of the \( i \)th flexible mode now computed according to 
\begin{equation}
\mathbf{X}_i\left(\boldsymbol{\rho}_i\right)=\left[\begin{array}{cccc}
2 \omega_i & 2 \zeta_i & 0 & 0 \\
0 & 2\omega_i & 0 & 0 \\
0 & 0 & \boldsymbol{\phi}_{r, i} \otimes \mathbf{I}_{n_{\mathrm{y}}} & \mathbf{I}_{n_{\mathrm{u}}} \otimes \boldsymbol{\phi}_{l, i}
\end{array}\right].
\end{equation}
\noindent
{
\subsection{Summary}
 The above derivations provide explicit parameter mappings and their associated Jacobians for both generally viscously damped and proportionally damped mechanical systems. These parameter mappings and their Jacobians are used to formulate and solve the IPEM problem, enabling the structured identification of modal models from the additive parameters estimated in the first stage and thereby completing the solution to the IPEM problem~\eqref{eq: IPEM estimation problem}. A complete overview of the proposed two-stage MIMO modal identification framework is provided in Procedure~\ref{alg: procedure}.}

\begin{algorithm}[t]
\caption{Two-stage approach for MIMO modal model estimation}
\label{alg: procedure}
\begin{algorithmic}[1]
\Require FRF dataset $\{G(j\omega_k)\}^N_{k=1}$, initialization $\{\omega^{\langle 0 \rangle}_i, \zeta^{\langle 0 \rangle}_i\}^{n_{\mathrm{flex}}}_{i=1}$, RIV tolerance $\epsilon_1$, IPEM tolerance $\epsilon_2$
\Statex \textbf{Stage 1: additive model estimation} 
\State  Construct $\bar A_i(s)$ from $\{\omega^{\langle 0 \rangle}_i, \zeta^{\langle 0 \rangle}_i\}^{n_{\mathrm{flex}}}_{i=1}$.
\State Solve the linear LS problem~\eqref{eq: convex problem} to obtain $\hat{\boldsymbol{\eta}}$ and construct $\boldsymbol{\beta}^{\langle 0 \rangle}$.
\State Solve the non-linear LS problem~\eqref{eq: FREQ - optimization problem} using the RIV Algorithm 1 to obtain  $\hat{\boldsymbol{\beta}}$.
\State Compute $\hat{\boldsymbol{\Sigma}}_{\hat{\boldsymbol{\beta}}}$ using \eqref{eq: cov}
\Statex \textbf{Stage 2: projection to modal model} 
\State Construct $\boldsymbol{\rho}^{\langle 0 \rangle}$ from $\hat{\boldsymbol{\beta}}$ using SVDs
\State Solve the IPEM problem \eqref{eq: IPEM estimation problem} using Algorithm 2 to obtain   $\hat{\boldsymbol{\rho}}$.
\Ensure Modal parameter vector estimate $\hat{\boldsymbol{\rho}}$
\end{algorithmic}
\end{algorithm}

\newpage
\section{Experimental validation}
Next, the two-stage identification procedure is experimentally validated on an experimental wafer-stage system. First, the experimental setup is introduced, followed by a discussion of the estimation process, and finally the results are presented.

\subsection{Experimental setup: experimental wafer-stage system} The considered experimental setup depicted in Figure \ref{fig: oat_body} is a prototype wafer stage system, which is designed as a lightweight mechanical structure to enable fast motions with high accelerations. The stage is magnetically levitated by passive gravity compensators positioned at its corners, establishing a mid-air equilibrium, and is actively controlled in six motion degrees of freedom at a sampling rate of 10 kHz, achieving sub-micrometer accuracy. The system contains 17 actuators, with 13 voice-coil actuators in the $ z $-direction and two Lorentz actuators for each $ x $- and $ y $-directions. {To measure the position and orientation of the stage, the system includes seven positioning sensors, consisting of four linear encoders for the \(z\)-direction, two capacitance sensors for the \(x\)-direction, and a single capacitance sensor for the \(y\)-direction.}  

\begin{figure}
        \centering
        \includegraphics[width=0.7\linewidth]{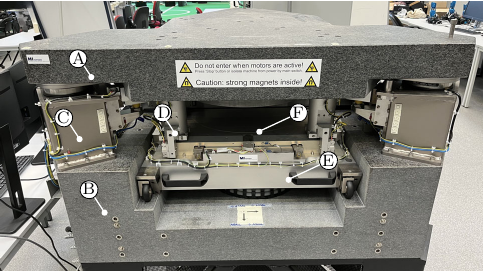}
        \caption{{Overview of the experimental wafer-stage setup. The sensory equipment is mounted on the metrology frame (A), which is isolated from the baseframe (B) through an air suspension system (C). The linear encoders for the out-of-plane measurements and the capacitance sensors for the in-plane measurements are mounted on the pillars (D), which are attached to the metrology frame. The actuation system is mounted on the force frame (E), and the stage (F) is isolated from the force frame by a set of gravity compensators.}}
        \label{fig: oat_overview 1}

        \vspace{2em}
        \includegraphics[width=0.6\linewidth]{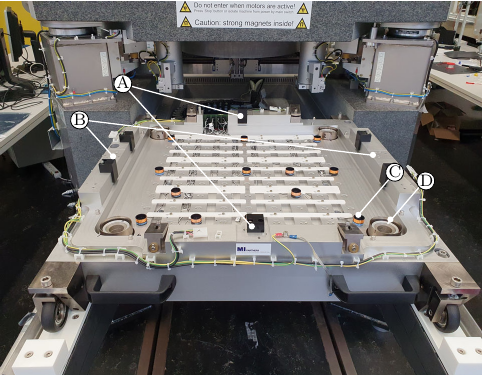}
        \caption{{Overview of the actuation system. The force frame is shown extracted from the machine, with the locations of the in-plane Lorentz motors for the \(x\)-direction shown in (A) and the actuator locations for the \(y\)-direction shown in (B). For the out-of-plane motion direction, 13 voice-coil actuators are used, with a single actuator shown in (C). Four passive gravity compensators are positioned at the corners (D) to compensate the gravitational load of the stage.}}
        \label{fig: oat_overview 2}
        
\end{figure}

{The complete input–output dynamics of the wafer-stage system are described by a \(7 \times 17\) transfer matrix. To simplify the presentation of the results, only the measurements in the out-of-plane direction are considered, namely translation along the \(z\) axis and rotations around the \(x\) and \(y\) axes. The corresponding input–output dynamics of this subset of actuators and sensors are described by a \(4 \times 13\) transfer matrix. A schematic overview of the actuator and sensor layout in the out-of-plane direction is provided in Figure~\ref{fig: oat_body}. In addition, in Figure \ref{fig: oat_overview 1} and Figure \ref{fig: oat_overview 2} a detailed overview is provided of the experimental setup.} The lightweight design of the stage results in high-order, lightly damped flexible dynamics that limit system performance, making accurate parametric models essential for achieving further performance improvements.

\subsection{Nonparametric modeling} 
\begin{figure}[b]
    \centering
    \begin{subfigure}{0.48\columnwidth}
        \centering
        \includegraphics[width=0.95\linewidth]{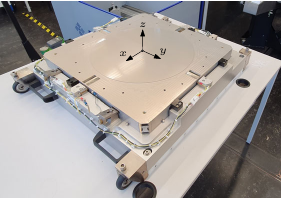}
    \end{subfigure}
        \hfill
     \centering
    \begin{subfigure}{0.48\columnwidth}
    \centering
    \includegraphics[width=0.75\columnwidth]{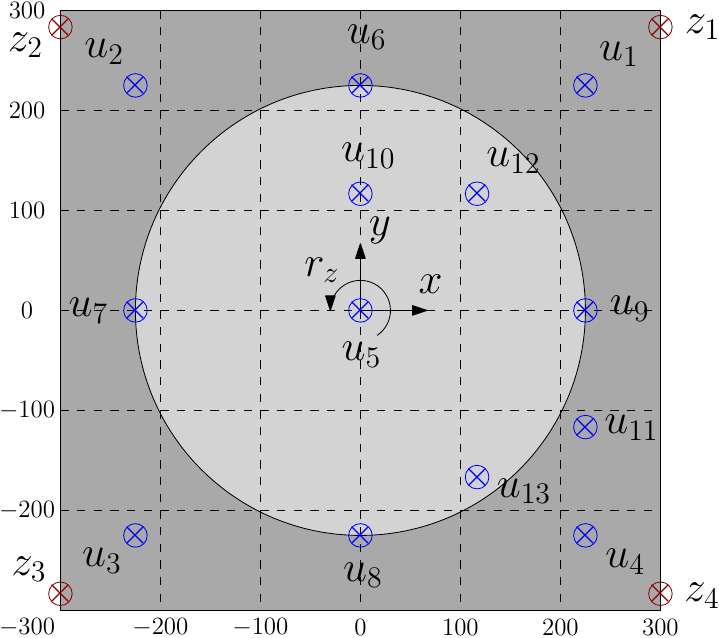}
    \end{subfigure}
    \caption{Wafer chuck with force frame extracted from the machine (left).  Schematic overview of the featured actuators $u_i$ and sensors $z_i$ in the out-of-plane direction (right).}
    \label{fig: oat_body}
\end{figure}

First, an FRF model of the wafer-stage system is estimated to serve as the dataset for subsequent parametric identification. The system consists of \( n_{\mathrm{y}}=4 \) outputs and \( n_{\mathrm{u}}=13 \) inputs in the out-of-plane direction, with actuator and sensor positions shown in Figure~\ref{fig: oat_body}. The FRF model of the $4 \times 13$ out-of-plane plant dynamics is obtained using the robust best linear approximation approach described in \cite[Chapter 3]{Pintelon2012SystemIdentification}. The experiments are performed in a closed-loop configuration since active control of the mid-air equilibrium is required to ensure stable operation. The plant FRF is derived using the indirect method, where the system is excited by $n_{\mathrm{u}} $ single-axis random-phase multisine signals with a flat amplitude spectrum. The multisine excitation includes $ 10 $ periods and $ 10 $ realizations, resulting in a plant FRF consisting of $ N = 4000 $ complex data points spanning a frequency range of 0.25 Hz to 2000 Hz. Frequency lines below 20 Hz are discarded during parametric estimation due to low FRF quality in this range.

\subsection{Model structure and model-order selection}
Next, the model structure and corresponding model orders used for parametric identification are specified. The general damping model structure is considered; however, it is noted that since the wafer-stage system is well described by a proportional damping assumption, the general formulation yields only a slight improvement in performance for this specific case. The wafer-stage system exhibits \( n_{\mathrm{rbm}} = 3 \) rigid-body modes in the out-of-plane direction, corresponding to translation along the \( z \)-axis and rotations around the \( x \)- and \( y \)-axes. 

\begin{figure}[b]
\centering
    \includegraphics[width = 0.8\linewidth]{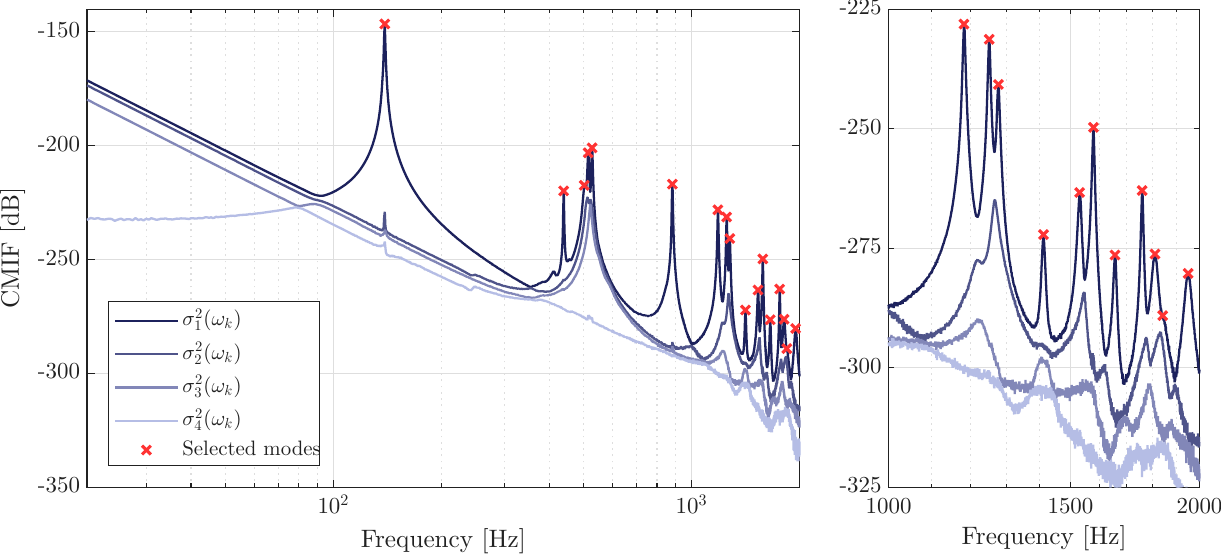}
  \caption{CMIF plot used for model order selection and initialization of the mode frequencies (left) with additional zoomed-in view (right), with $\sigma_i(\omega_k)$ denoting the $i$th singular value of the measured FRF dataset at frequency $k$. The frequency locations of the flexible modes are indicated by peaks in the singular-value plot, with multiplicity determined by the number of singular values peaking at each frequency.}
  \label{fig:cmif}
\end{figure}

The number of flexible modes \( n_{\mathrm{flex}} \) is determined using the CMIF~\cite{Shih1988ComplexEstimation}, which is computed from the squared singular values of the FRF matrix at each frequency. The corresponding modal frequencies are identified from the peak locations in the CMIF, while mode multiplicity is inferred from the number of singular values peaking at the same frequency.  From the CMIF, shown in Figure~\ref{fig:cmif}, \( n_{\mathrm{flex}} = 17 \) distinct flexible modes are identified. Thus, the used model structure consists of \( n_{\mathrm{rbm}} = 3 \) rigid-body modes, \( n_{\mathrm{flex}} = 17 \) flexible modes, and an additional DC gain to capture the effect of higher-order dynamics, which results in a 40th-order parametric model.

\begin{figure}[b!]
    \centering
    \includegraphics[scale = 0.9]{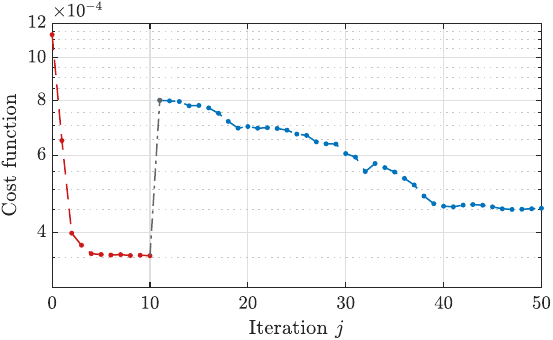}
    \caption{Cost-function evolution over iterations. 
    Stage~1 (\(j \leq 10\)) corresponds to the \tikzmarkline{MatlabRed} RIV algorithm, while Stage~2 (\(j > 10\)) applies the \tikzmarkline{MatlabBlue} IPEM algorithm. }

    \label{fig: cost function}

  \centering
  \includegraphics[scale = 1.05]{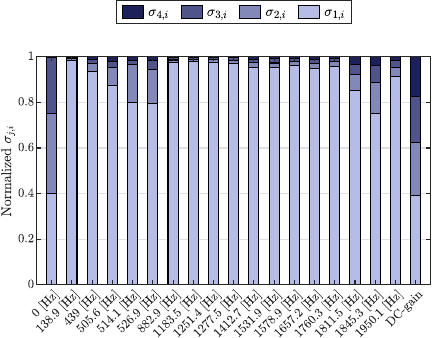}
  \caption{Normalized singular values of the estimated residue matrices after the first stage ($j=10$). The mode at $f=0$ [Hz] shows 3 dominant directions, consistent with the expected $n_{\mathrm{rbm}}=3$, whereas the flexible modes mostly show a single dominant direction. The DC-gain is full-rank as it approximates the higher-order modes. Since the flexible modes are not strictly rank-one, they cannot be represented exactly by a modal model, motivating the need for further refinement. }
  \label{fig: svd additive model}
\end{figure}

\subsection{Weighting filter design}
To balance the contribution of different frequency regions in the identification, a weighting filter is introduced. An element-wise inverse plant magnitude weighting is selected, given by 
\begin{equation}
    \mathbf{W}(\omega_k) = \operatorname{diag}\Bigl(\operatorname{vec}\left(|\mathbf{G}(\omega_k)|\right)\Bigr)^{-1}.
\end{equation}
The inverse plant magnitude weighting effectively transforms the matrix residual \eqref{eq: FREQ - residual} from absolute to relative error criterion. This prevents overemphasizing frequencies with a large magnitude, which can dominate the estimation process, especially for systems containing integrator dynamics. 

\subsection{Initialization}
Due to the non-convexity of the cost function in \eqref{eq: FREQ - optimization problem}, accurate initialization of the estimator \eqref{eq: FREQ - estimator} is essential. To initialize the stage one estimation problem, initial pole locations must be provided, which enable the numerators to be computed using the convex LS problem \eqref{eq: convex problem}. The pole locations are constructed from initial estimates \(\omega_i\) and \(\zeta_i\), where the natural frequencies \(\omega_i\) are set to the CMIF peak frequencies, and damping ratios \(\zeta_i\) are uniformly assigned a value of 0.01, representative of lightly damped flexible modes. 

\begin{figure}[t!]
    \centering    
    \includegraphics{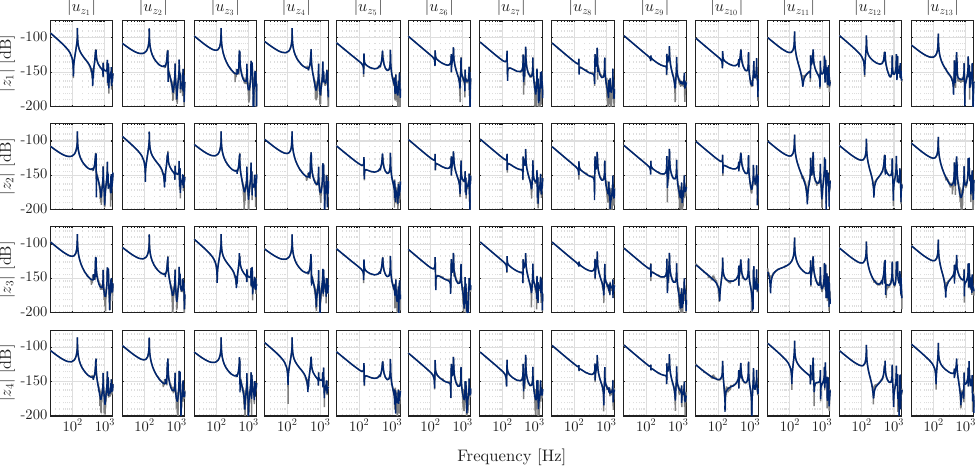}
    \caption{Element-wise Bode magnitude plot of the the FRF measurement $\mathbf{G}(\omega_k)$ \tikzline{MatlabGray50} and the estimated parametric modal model $\hat{\mathbf{P}}_m(j\omega_k)$ \tikzline{MatlabBlue}.}
    \label{fig: oat_fit}
    \vspace{0.5em}
    
    \centering
    % First Column
    \begin{minipage}{0.31\textwidth}
        \centering
        \includegraphics[width=\linewidth]{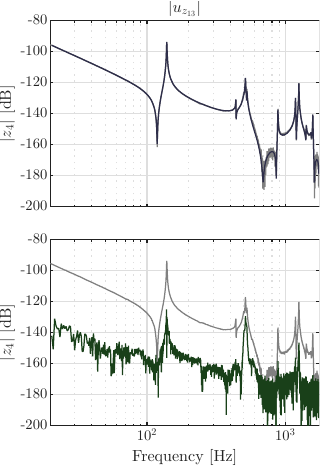}\\
    \end{minipage}
    \hfill
    % Second Column
    \begin{minipage}{0.31\textwidth}
        \centering
        \includegraphics[width=\linewidth]{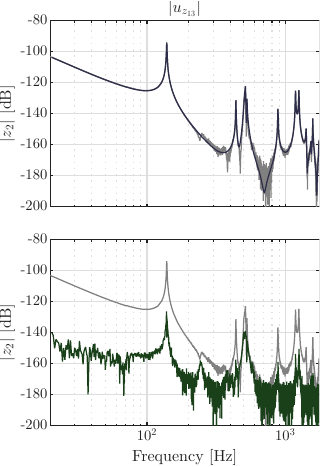}\\
    \end{minipage}
    \hfill
    % Third Column
    \begin{minipage}{0.31\textwidth}
        \centering
        \includegraphics[width=\linewidth]{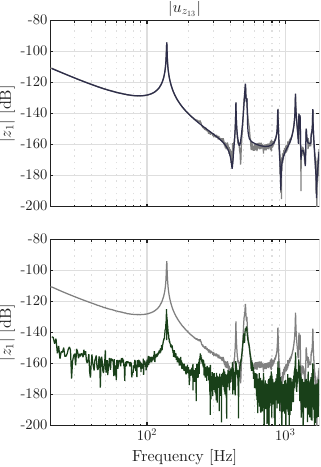}\\
    \end{minipage}

    \caption{Bode magnitude plot (top) with the corresponding residual (bottom). Shown are the FRF measurement $\mathbf{G}(\omega_k)$ \tikzline{MatlabGray50}, the estimated parametric modal model $\hat{\mathbf{P}}_m(j\omega_k)$ \tikzline{MatlabBlue}, and the residual ${\mathbf{E}}_m(j\omega_k)$ \tikzline{MatlabGreen}.}
    \label{fig: oat_fit_single}
\end{figure}

\begin{figure}[p]
    \centering    
    \includegraphics[width=\linewidth]{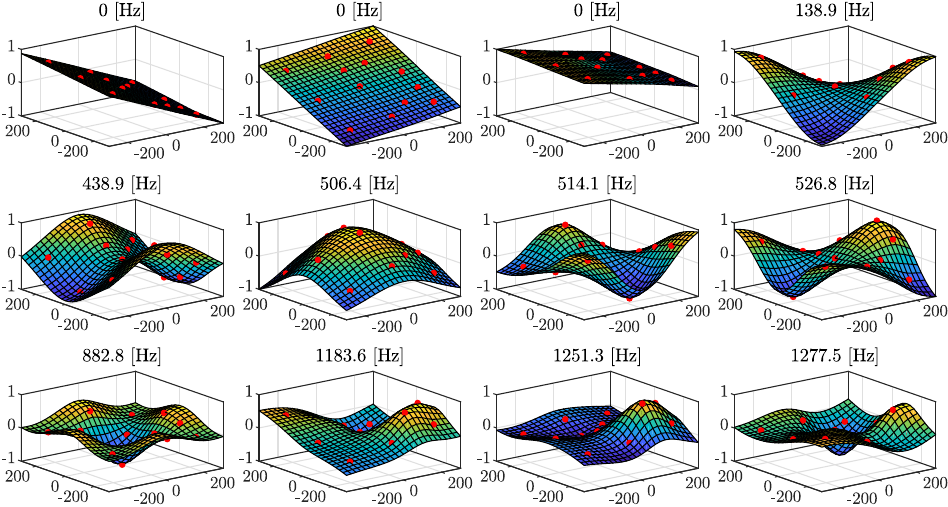}
    \caption{Experimentally identified mode shapes of the experimental wafer-stage system (actuator side) for the first 12 modes. The shapes are interpolated using the thin-plate spline method introduced in \cite{Voorhoeve2021IdentifyingStage}. The first 3 modes correspond to rigid-body modes, while the remaining 9 are flexible modes.}
    \label{fig: oat modeshapes}
    
    \vspace{2em}

    \includegraphics[width=\linewidth]{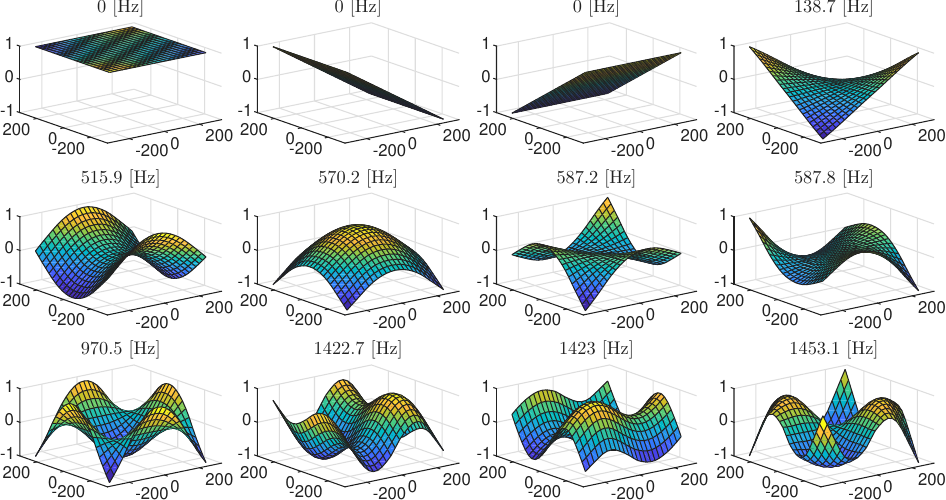}
    \caption{Theoretical mode shapes obtained from the FEM model of the wafer-stage system for the first 12 nontrivial out-of-plane modes. The shapes are interpolated onto a regular grid using the same thin-plate spline procedure as used for the experimentally identified modes. The first 3 modes represent rigid-body motion, while the remaining 9 correspond to flexible structural modes.}
    \label{fig: oat fem modeshapes}
    \vspace{0.5em}
\end{figure}

\subsection{Results}
The two-stage parametric identification procedure is applied to estimate a parametric model of the wafer-stage system. {In the first stage, an additive model is estimated by solving (21) using the RIV approach. In the second stage, the resulting estimate is projected onto a modal model by solving the IPEM problem (22), which is implemented using available gradient-based numerical solvers in MATLAB.} In Figure \ref{fig: cost function}, the evolution of the cost function \eqref{eq: FREQ - optimization problem} is shown. It is emphasized that the RIV method does not guarantee monotonic convergence in general but typically demonstrates fast convergence to a locally optimal solution. Figure \ref{fig: svd additive model} shows the singular value distribution of the estimated residue matrices after the first stage ($j=10$). This distribution reflects the extent to which the model approximates an ideal modal form. In an exact modal representation of flexible modes, the residue matrices in \eqref{eq: resid prop} must be exactly rank one. However, the estimated residue matrices contain additional nonzero singular values, with the largest discrepancy for the modes around 500 Hz and 1800 Hz, motivating the need for further refinement. The rigid body mode shows three dominant singular values, consistent with $n_{\mathrm{rbm}} = 3$, and since the DC gain must approximate many higher‐order modes, its residue matrix is expected to be full rank.

Next, the second-stage IPEM problem defined in \eqref{eq: GN iterations} is solved. First, an initial modal model is computed using the SVD-based decomposition, which is subsequently refined through the projection step. The frequency response of the identified modal model is shown in Figure \ref{fig: oat_fit}, together with the FRF measurement used in the estimation. Furthermore, Figure \ref{fig: oat_fit_single} shows the frequency response of individual plant entries, along with the corresponding model error. These figures shows a good agreement between the modal model and the FRF over the complete frequency range. In the frequency range around 500 Hz, the modeling error is slightly larger, suggesting either the presence of higher mode multiplicities or of unmodeled dynamics not captured by the modal-model structure.

Finally, Figure~\ref{fig: oat modeshapes} visualizes the estimated mode-shape vectors at the actuator side. Real-mode shape-vectors are provided corresponding to the proportionally damped model structure, which are interpolated using thin-plate splines (see~\cite{Voorhoeve2021IdentifyingStage} for details). {In addition, Figure~\ref{fig: oat fem modeshapes} shows the theoretical mode shapes obtained from a Finite-Element Method (FEM) analysis of the wafer-stage system, enabling a qualitative comparison between the estimated and modeled dynamics. Details of the FEM model are provided in Appendix~B. The first three modes correspond to rigid-body motion, while the remaining modes represent flexible dynamics. The first nine flexible modes show a clear correspondence between the experimental and FEM mode shapes, whereas the higher modes show deviations due to the limited spatial resolution of the available actuators and sensors.} These mode shapes provide valuable insight into the spatial deformation behavior of the wafer-stage system and support both controller design and system analysis.

\section{Conclusions}
This paper introduced a parametric identification method for complex, MIMO mechanical systems in a modal model structure, that accommodates both general‐viscous and proportional damping formulations. By decomposing the structured identification problem into two sequential estimation stages, the proposed framework enables computationally efficient estimation of modal models for high‐dimensional systems.  Experimental validation on a prototype wafer‐stage system with 13 actuators and 4 sensors produced an accurate 40th‐order modal model, comprising three rigid‐body modes and 17 flexible modes, that closely matches the measured FRF. These results demonstrate the method’s effectiveness and robustness in modeling the dynamics of high‐dimensional industrial systems. 

\appendix

\section{Modal representations of mechanical systems}
This Appendix derives the modal form for the general‐viscous damping case in \eqref{eq: modal transfer first order} and for the proportional damping case in \eqref{eq: modal transfer second order propportional}. Finally, a minimal and real state-space representation is provided describing both forms. 

\subsection{General-viscously damped systems}
First, the general case is considered, where the damping matrix $\mathbf{D}$ can be an arbitrary matrix. The system of equations in \eqref{eq:  eom nodal coordinates} are written in the descriptor state-space form
\begin{equation}
\begin{aligned} \label{eq: descriptor system}
\mathbf{E}\dot{\mathbf{x}}(t) &= \mathbf{A}\mathbf{x}(t) + \mathbf{B} \mathbf{u}(t),\\
\mathbf{y}(t) &= \mathbf{C} \mathbf{x}(t),
\end{aligned}
\end{equation}
with the state \(\mathbf{x} = [\mathbf{q}^\top, \dot{\mathbf{q}}^\top]^\top\) and with the system matrices defined as
\begin{equation}
\begin{aligned} \label{eq: descriptor system - matrices}
\mathbf{E} = \begin{bmatrix}
\mathbf{D} & \mathbf{M} \\
\mathbf{M} & \mathbf{0}
\end{bmatrix},\quad 
\mathbf{A} = \begin{bmatrix}
-\mathbf{K} & \mathbf{0} \\
\mathbf{0} & \mathbf{M}
\end{bmatrix},\quad 
\mathbf{B} = \begin{bmatrix}
\mathbf{F} \\
\mathbf{0}
\end{bmatrix},\quad
\mathbf{C} = \begin{bmatrix}
\mathbf{Q} & \mathbf{0}
\end{bmatrix}.
\end{aligned}
\end{equation}
A coordinate transformation to a modal form is obtained by solving the generalized eigenvalue problem
\begin{align}
(\mathbf{A} - \lambda_i \mathbf{E}) \mathbf{v}_i = \mathbf{0}, \quad i = \{1 \dots 2n\},
\end{align}
where \(\mathbf{v}_i \in \mathbb{C}^{2n}\) is the complex mode-shape vector, normalized according to \(\mathbf{v}_i^\top \mathbf{E} \mathbf{v}_i = 1\), and \(\lambda_i \in \mathbb{C}\) is the corresponding eigenvalue given in \eqref{eq: eigenvalue}. The state relation described by the matrices \eqref{eq: descriptor system - matrices} enables the modal matrix \(\mathbf{V} = [\mathbf{v}_1, \ldots, \mathbf{v}_{2n}]\) to be partitioned as \(\mathbf{V} = \left[ \mathbf{U}^\top, \boldsymbol{\Lambda}\mathbf{U}^\top \right]^\top\), with \(\mathbf{U} \in \mathbb{C}^{n \times 2n}\) containing the unique mode shape information, and the matrix \(\boldsymbol{\Lambda} = \operatorname{diag}([\lambda_1, \ldots, \lambda_{2n}])\) contains the eigenvalues along the main diagonal. The descriptor system \eqref{eq: descriptor system} is transformed to a modal system with states \(\mathbf{x}_m \in \mathbb{R}^{2n}\) by substituting \(\mathbf{x}(t) = \mathbf{V} \mathbf{x}_m(t)\) followed by premultiplication with \(\mathbf{V}^\top\) to obtain
\begin{equation} \label{eq: modal state-space complex}
\begin{aligned} 
\dot{\mathbf{x}}_m(t) &= \boldsymbol{\Lambda} \mathbf{x}_m(t) + \boldsymbol{\Psi}_r^\top \mathbf{u}(t),\\
\mathbf{y}(t) &= \boldsymbol{\Psi}_l \mathbf{x}_m(t),
\end{aligned}
\end{equation}
where \(\boldsymbol{\Psi}_r^\top = \mathbf{U}^\top \mathbf{F} = [\boldsymbol{\psi}_{r,1}, \dots, \boldsymbol{\psi}_{r,2n}]^\top\) is the modal input matrix, \(\boldsymbol{\Psi}_l = \mathbf{Q} \mathbf{U} = [\boldsymbol{\psi}_{l,1}, \dots, \boldsymbol{\psi}_{l,2n}]\) is the modal output matrix, and \(\boldsymbol{\psi}_{l,i} \in \mathbb{C}^{n_{\mathrm{y}}}, \boldsymbol{\psi}_{r,i} \in \mathbb{C}^{n_{\mathrm{u}}}\) are the complex left and right mode-shape vectors, respectively. The transfer matrix representation of \eqref{eq: modal state-space complex} is obtained using the Laplace transform, yielding
\begin{align} \label{eq: transfer matrix}
\mathbf{P}_m(s) &= \boldsymbol{\Psi}_l  \left(s \mathbf{I}_{2n} - \boldsymbol{\Lambda} \right)^{-1} \boldsymbol{\Psi}_r^\top,
\end{align}
which can equivalently be written as a first-order partial-fraction expansion 
\begin{align}
\mathbf{P}_m(s) &= \sum_{i=1}^{2n}\frac{\boldsymbol{\psi}_{l,i} \boldsymbol{\psi}_{r,i}^\top}{s - \lambda_i}.
\end{align}
For real-valued state matrices in \eqref{eq: descriptor system - matrices}, the eigenvalues \(\lambda_i\) and mode-shape vectors \(\mathbf{v}_i\) appear as complex conjugate pairs. This property enables the first-order form to be partitioned according to
\begin{align} \label{eq: GD - first-order}
\mathbf{P}_m(s) &= \sum_{i=1}^{n}\frac{\boldsymbol{\psi}_{l,i} \boldsymbol{\psi}_{r,i}^\top}{s - \lambda_i} + \frac{\bar{\boldsymbol{\psi}}_{l,i} \bar{\boldsymbol{\psi}}_{r,i}^\top}{s - \bar{\lambda}_i},
\end{align}
which after introducing the partitioning of rigid-body and flexible modes, yields the form in \eqref{eq: modal transfer first order}, concluding the derivation.

\subsection{Proportionally-damped systems}
For proportionally damped systems, the damping matrix $\mathbf{D}$ is such that the damped system shares the same modal basis as the undamped system. Hence, a coordinate transform to modal coordinates can be obtained by solving the generalized eigenvalue problem for the undamped system
\begin{equation}
    (\mathbf{K} - \omega_i^2 \mathbf{M}) \boldsymbol{\phi}_i = \mathbf{0}, \quad i = \{1,\ldots,n\},
\end{equation}
where \(\boldsymbol{\phi}_i\in\mathbb{R}^n\) are the mode-shape vectors, which are normalized with respect to the mass matrix according to \(\boldsymbol{\phi}_i^\top \mathbf{M} \boldsymbol{\phi}_i = 1\). By substituting \(\mathbf{q}(t) = \boldsymbol{\Phi} \mathbf{q}_m(t)\) in \eqref{eq: eom nodal coordinates}, followed by pre-multiplication with \(\boldsymbol{\Phi}^\top\), the equations of motion in modal coordinates are obtained
\begin{equation} \label{eq: mod eom}
\begin{aligned}
&\ddot{\mathbf{q}}_m(t) + 2 \mathbf{Z} \boldsymbol{\Omega} \dot{\mathbf{q}}_m(t) + \boldsymbol{\Omega}^2 \mathbf{q}_m(t) = \boldsymbol{\Phi}_r^\top \mathbf{u}(t),\\
&\mathbf{y}(t) = \boldsymbol{\Phi}_l \mathbf{q}_m(t),
\end{aligned}
\end{equation}
with \(\boldsymbol{\Omega} = \operatorname{diag}([\omega_1, \ldots, \omega_n])\), \(\mathbf{Z} = \operatorname{diag}([\zeta_1, \ldots, \zeta_n])\) and where \(\boldsymbol{\Phi}_r^\top = \boldsymbol{\Phi}^\top \mathbf{F} = [\boldsymbol{\phi}_{r,1},\ldots, \boldsymbol{\phi}_{r,n}]\) is the modal input matrix and \(\boldsymbol{\Phi}_l = \mathbf{Q}\boldsymbol{\Phi} = [\boldsymbol{\phi}_{l,1},\ldots, \boldsymbol{\phi}_{l,n}]\) is the modal output matrix. The transfer matrix representation is obtained through the Laplace transform
\begin{align} \label{eq: transfer matrix}
\mathbf{P}_m(s) &= \boldsymbol{\Phi}_l  \left(s \mathbf{I}_{2n} + 2 \mathbf{Z} \boldsymbol{\Omega} + \boldsymbol{\Omega}^2 \right)^{-1} \boldsymbol{\Phi}_r^\top,
\end{align}
which can equivalently be written as a second-order partial-fraction expansion
\begin{align} \label{eq: p}
\mathbf{P}_m(s) &=  \sum_{i=1}^{n}  \frac{\boldsymbol{\phi}_{l,i} \boldsymbol{\phi}_{r,i}^\top}{s^2 + 2 \zeta_i \omega_i s + {\omega}_i^2}, 
\end{align}
which corresponds to the provided form in \eqref{eq: modal transfer second order propportional}. Note that \eqref{eq: p} is a special case of the general form and, under an appropriate scaling of the mode-shape vectors, directly follows from \eqref{eq: GD - first-order} (see \cite{DeKraker2004ADynamics} for a proof).

\subsection{State‐space representation}
For control applications it is often preferable to work with a real‐valued, minimal state‐space model. A real realization of \eqref{eq: modal state-space complex} is obtained via a similarity transformation with the transformation matrix  
\begin{equation} \label{eq: complex to real map}
\mathbf{T} =  \begin{bmatrix}
\bar{\boldsymbol{\Lambda}}_n & -\mathbf{I}_n \\
\boldsymbol{\Lambda}_n & -\mathbf{I}_n
\end{bmatrix},
\end{equation}
where \(\boldsymbol{\Lambda}_n = \operatorname{diag}([\lambda_1,\ldots, \lambda_n])\) contains the \(n\) unique eigenvalues along the diagonal. Applying the similarity transform yields the transformed system
\begin{equation} \label{eq: modal state-space system real}
\left[\begin{array}{c}
\dot{\mathbf{x}}_m(t) \\
 \mathbf{y}(t)
\end{array}\right] = 
\left[\begin{array}{cc|c}
\mathbf{0} & \mathbf{I}_n & \tilde{\boldsymbol{\Phi}}_r^{\top} \\
-\boldsymbol{\Omega}^2 & -2\boldsymbol{\mathbf{Z}}\boldsymbol{\Omega} & \boldsymbol{\Phi}_r^{\top} \\
\hline  \boldsymbol{\Phi}_l & \tilde{\boldsymbol{\Phi}}_l & \mathbf{0}
\end{array}\right] 
\left[\begin{array}{c}
\mathbf{x}_m(t) \\
\mathbf{u}(t)
\end{array}\right],
\end{equation}
where \(\boldsymbol{\Phi}_r, \tilde{\boldsymbol{\Phi}}_r \in \mathbb{R}^{n_{\mathrm{u}} \times n}\), \(\boldsymbol{\Phi}_l, \tilde{\boldsymbol{\Phi}}_l \in \mathbb{R}^{n_{\mathrm{y}} \times n}\) are the real modal input and output matrices, respectively. Furthermore, the state-space form of \eqref{eq: mod eom} directly corresponds to \eqref{eq: modal state-space system real} with $\tilde{\boldsymbol{\Phi}}_r,\tilde{\boldsymbol{\Phi}}_l = \mathbf{0}$. Hence, the key difference between modal forms is that the general damping model introduces an additional coupling between the position and velocity of each modal state whereas the proportional damping model results in each modal state depending solely on position. The state-space form  \eqref{eq: modal state-space system real} provides a real and minimal representation of the mechanical system described by \eqref{eq: eom nodal coordinates} for both damping models and can be constructed directly using the parameters estimated in the two-stage identification procedure.

\newpage 
\section{FEM model OAT}
\begin{figure}[t!]
\centering
    \includegraphics[width = 0.6\linewidth]{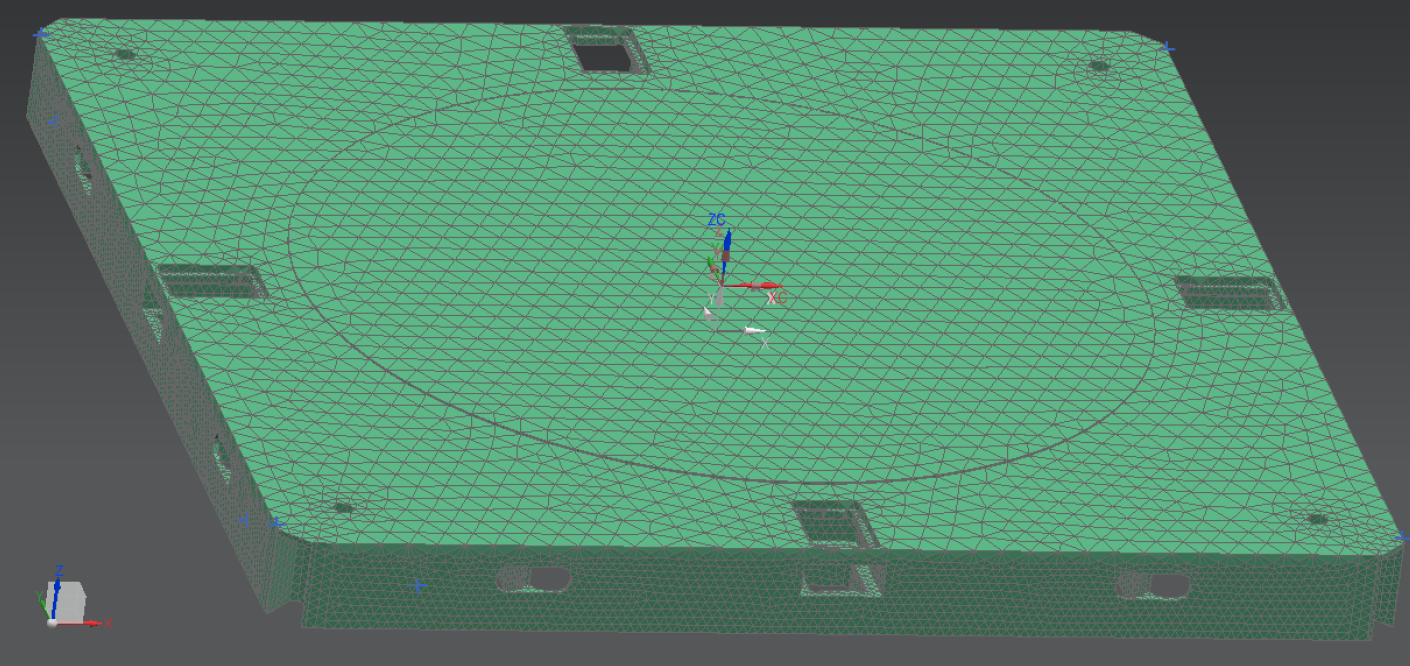}
\caption{FEM model of the wafer-stage system.}
  \label{fig: fem model}
\end{figure}

The FEM model is derived from a Siemens NX geometry of the wafer stage and discretized using tetrahedral elements, resulting in a mesh with 85 262 nodes and 259 185 elements. The wafer stage is modeled as a thin rectangular structure with a width of 0.6 [m],  length 0.6 [m] and a height of 45 [mm], which for the considered material properties leads to a mass of approximately 8.6 [kg]. The FEM model is used to compute the theoretical mode shapes and eigenfrequencies of the experimental wafer-stage system, which serve as a reference for validating the experimentally identified modes. The FEM model can be seen in Figure \ref{fig: fem model}.

\bibliographystyle{abbrv} % Abbreviated author names
% \bibliography{References} % Use the name of your .bib file without the extension
 \bibliography{root_arxiv_submission.bbl} % Use the name of your .bib file without the extension
\end{document}